\documentclass[preprint2]{aastex}

\catcode`\@=11
\newcommand{\gapprox}{\mathrel{\mathpalette\@versim>}}
\newcommand{\lapprox}{\mathrel{\mathpalette\@versim<}}
\newcommand{\propapprox}{\mathrel{\mathpalette\@versim\propto}}
\newcommand{\@versim}[2]
  {\lower3.1truept\vbox{\baselineskip0pt\lineskip0.5truept
\ialign{$\m@th#1\hfil##\hfil$\crcr#2\crcr\sim\crcr}}}
\catcode`\@=12

\shorttitle{PSR J0737$-$3039B: A PROBE OF RADIO PULSAR EMISSION HEIGHTS}
\shortauthors{PERERA ET AL.}

\begin{document}

\title{PSR J0737$-$3039B: A probe of radio pulsar emission heights}

\author{B.~B.~P.~Perera,\altaffilmark{1} D.~Lomiashvili,\altaffilmark{2} K.~N.~Gourgouliatos,\altaffilmark{2}
 M.~A.~McLaughlin,\altaffilmark{1,3} M.~Lyutikov\altaffilmark{2}}

\altaffiltext{1}{Department of Physics, West Virginia University,
Morgantown, WV 26506, USA.}
\altaffiltext{2}{Department of Physics, Purdue University, West Lafayette, IN 47907, USA.}
\altaffiltext{3}{Also an adjunct astronomer at the National Radio Astronomy Observatory, Green Bank,
WV 24944,USA.}

\vskip 1 truein

\newpage

\begin{abstract}

In the double pulsar system PSR J0737$-$3039A/B the strong wind produced by pulsar A distorts the magnetosphere of pulsar B. The influence of these distortions on the  orbital-dependent emission properties of pulsar B can be used to determine the location of the coherent radio emission generation region in the pulsar magnetosphere. Using a model of the wind-distorted magnetosphere of pulsar B and the well defined geometrical parameters of the system, we determine the minimum emission height to be $\sim 20 R_{NS}$ in the two bright orbital longitude regions.
We can determine the maximum emission height by accounting for the amount of deflection of the polar field line with respect to the magnetic axis using the analytical magnetic reconnection model of Dungey and the semi-empirical numerical model of Tsyganenko. Both of these models estimate the maximum emission height to be $\sim 2500 R_{NS}$.  The minimum and maximum emission heights we calculate are consistent with those estimated for normal isolated pulsars.

\end{abstract}


\section{Introduction}
\label{intro}

The mechanism of pulsar radio emission and its origin within the pulsar magnetosphere are not well understood \citep[see, e.g.,][]{hre+09}. In general, it is thought to be due to coherent radiation from relativistic plasma streaming along open magnetic field lines.
Radio emission height estimates can constrain the emission mechanism  to some extent. In normal isolated pulsars, radio emission heights have been estimated from their emission geometry inferred from radio polarization combined with the rotating vector model \citep{rc69a} and the pulse profile widths \citep{gk93, kg97, kg03a}. \citet{gg01} and \citet{drh04} have also proposed a phase-shift method to determine the emission height. In general, these methods show that core component emission originates very close to the surface of the neutron star (NS), but the conal components come from well above the surface \citep{ran90,mr01}. However, these techniques are limited in that we observe only a small section of the magnetosphere of these isolated pulsars due to an unchanging line-of-sight.

PSR J0737$-$3039A/B is a unique binary system which provides an excellent opportunity to study different emission regions due to relativistic spin precession, allowing us to observe different portions of the magnetosphere. We can also explore magnetospheric distortion,
 which affects  the observed emission pattern. The two neutron stars of this system orbit each other in a 2.4-hr orbit; this is the only known pulsar binary system in which both neutron stars have been detectable as radio pulsars \citep{bdp+03,lbk+04}. The first-born recycled pulsar $-$ hereafter A $-$ has a spin period of $23$ ms  and the second born pulsar $-$ hereafter B $-$ has a spin period of 2.8~s. The pulse profile of A has been stable since its discovery, but that of B has dramatically evolved through five years of observing, culminating in its radio disappearance in 2008 March \citep{pmk+10}.

Due to the unstable features of B's pulse profiles, both on long timescales and within a single orbit, it is challenging to understand the emission geometry and the emission mechanism. By fitting a model to eclipses of A due to absorption in the magnetosphere of B, \citet{bkk+08} constrained the geometrical parameters to be $\alpha = 70.9(4)\degr$ and $\theta = 130.0(4)\degr$ \footnote{Here, and throughout the paper, the number in parentheses is the  1-$\sigma$ uncertainty in the last quoted digit.}.  Here $\alpha$ is the misalignment of the magnetic axis with respect to the spin axis and $\theta$ is the colatitude of the spin axis. Unfortunately, B shows very little radio polarization, making it impossible to constrain the geometry from polarization measurements \citep{drb+04}. \citet{bkk+08} also constrained the precessional phase of the spin axis, measured from our line-of-sight, to be $\phi_{prec} = 51.2(8)\degr$ at an epoch of 2006 May 2 (i.e. MJD 53857) and found that this phase is changing with time at a rate of $4.8(7)\degr$~yr$^{-1}$. This is consistent with the rate of $5.061(2)\degr$~yr$^{-1}$ predicted by general relativity \citep{bo75b}. Recently, \citet{pmk+10} analyzed the  pulse profile evolution of B and independently determined the above angles using a simple model based on geodetic spin precession as proposed by \citet{cw08}. According to this model, the beam must be elliptical and horse-shoe shaped in order to explain the observed single- to double-peak pulse profile evolution and the disappearance of radio emission. The estimations of the above angles in this model are consistent with those predicted by \citet{bkk+08}, within the 2-$\sigma$ errors. From these studies, we believe that the geometrical parameters of B are well known. We can therefore use these values in this paper to determine the emission geometry.

In this paper, we also explore the distortion of the magnetosphere of  B.
The almost edge-on orbital plane of the system, with inclination angle of $88.7\degr$ \citep{ksm+06}, allows us to observe the eclipses of A with a duration of about 30 seconds. By considering the relative transverse velocities of the two pulsars, 660~km/s \citep{lbk+04}, the estimated size of B's magnetosphere is about $10\%$ of its light cylinder radius of $\sim 1.3\times10^{10}$~cm . This implies that the wind of A compresses the magnetosphere of B and disturbs its polar cap \citep{lyu04}. This is due to the small separation of the pulsars ($\sim 9\times 10^{10}$ cm or $2.9$ lt-s) and the large spin-down luminosity of A ($5.8\times10^{33}$~ergs$^{-1}$) compared to B ($1.6\times10^{30}$~ergs$^{-1}$). This is analogous to the distortion of the Earth's magnetosphere due to the Solar wind. This interaction also results in an orbital modulation of B; we detect bright emission from the pulsar only in two orbital phase regions of $185\degr-235\degr$ (hereafter BP1) and $265\degr-305\degr$ (hereafter BP2), and detect weak emission at phases of $340\degr-30\degr$ and $80\degr-130\degr$ (Here, and throughout the paper, orbital phases are measured from the ascending node). \citet{lyu05} claims that the pulsar has the same intrinsic radio intensity throughout the orbit and that the orbital modulation is due to the deflection of the magnetic polar field line with respect to the line-of-sight because of the influence of A.

The wind interaction with the magnetosphere of B produces a bow shock between A and B; this is likely the boundary of the magnetosphere of B. The shape of this boundary  depends on the orientation of the magnetic axis of B.
\citet{lyu04} constrained the stand-off distance, or the distance from  B to the vertex of the bow shock, to be $3.5\times10^{9}$~cm if the bow shock interface is a perfect resistor and $4\times10^{9}$~cm if it is partially resistive.
These estimates inferred that the magnetosphere of B is located deep within its light cylinder and the open and closed field lines have a more complicated structure than that of an isolated pulsar. Since the wind-interaction boundary model is very important to study the emission geometry of B, we derive it again in this paper with some improvements. This model describes the shape of the boundary for any orientation of the magnetic axis and allows us to model the open and closed field line structure more accurately.

This results in a method to use the derived field line structure to estimate the radio emission heights of B.
Since the bow shock boundary is located deep inside the light cylinder, the correction due to rotation on the static dipole field, which introduced in retarded dipole field, is small \citep[for retarded dipole field, refer appendix A of][]{dh04}. Therefore we assume non-rotating dipole field throughout the model. In our method, we assume that the emission comes from the direction tangential to the local field lines. We also assume that the emission comes from above the polar cap region, consistent with the narrow single- and double-peaked radio profiles.

The distortion of the Earth's magnetosphere due to the solar wind has been studied using a large number of satellite observations and these data have been extensively modeled. Since the wind of A distorts B's magnetosphere in the same way that the solar wind does the Earth's, some models for the Earth's  magnetic field line structure can be used  to study the distortion of the magnetosphere of B and to determine the regions of radio emission. We use the \citet{dun61} planetary magnetosphere model and the \citet{tsy02a,tsy02b} Earth magnetosphere models to set an upper limit on the radio emission height of B.

We present our observational data in section~\ref{obs} and discuss observed mean pulse profiles. In section \ref{bmodel}, we explain the boundary model which describes the shape of the bow shock. Then we trace the dipole field lines and transform them from the co-rotating frame of the neutron star to the orbit-fixed frame. In order to derive the required angular radius of the elliptical beam for the emission height calculation, we re-analyze the beam with a different geometrical framework in section~\ref{remodel}. In section \ref{aceh}, we present the method which we use to estimate the minimum emission height and our results. In section \ref{uplimit}, we describe the maximum emission height calculation using two different magnetosphere models. Finally in section \ref{dis}, we discuss our results and compare them with predicted emission heights for other pulsars. We also discuss height estimations from other methods, concluding that these are not applicable to pulsar B.

\section{Observations and pulse profiles}
\label{obs}

We observed J$0737-3039$B with the 100~m Green Bank Telescope (GBT) in West Virginia since 2003 December 24 at multiple frequencies. Since 820~MHz is the most common, we use only those data in this analysis.
This is the same data set which we reported in \citet{pmk+10}. However, we use better time resolution pulse profiles in this analysis compared to the previous paper. The data were taken using the GBT spectrometer SPIGOT with sampling time of 81.92~$\mu$s until 2009 January. After 2009 January, the spectrometer GUPPI was used with a sampling time of 61.44~$\mu$s. All the data were dedispersed and folded using the pulsar analysis package SIGPROC, assuming a dispersion measure of 48.914~cm$^{-3}$~pc \citep{lbk+04}. The ephemeris of \citet{lbk+04} was used until 2006 and since then we have used the ephemeris of \citet{ksm+06} to form mean pulse profiles.

\begin{figure*}
\epsscale{2.0}
\plotone{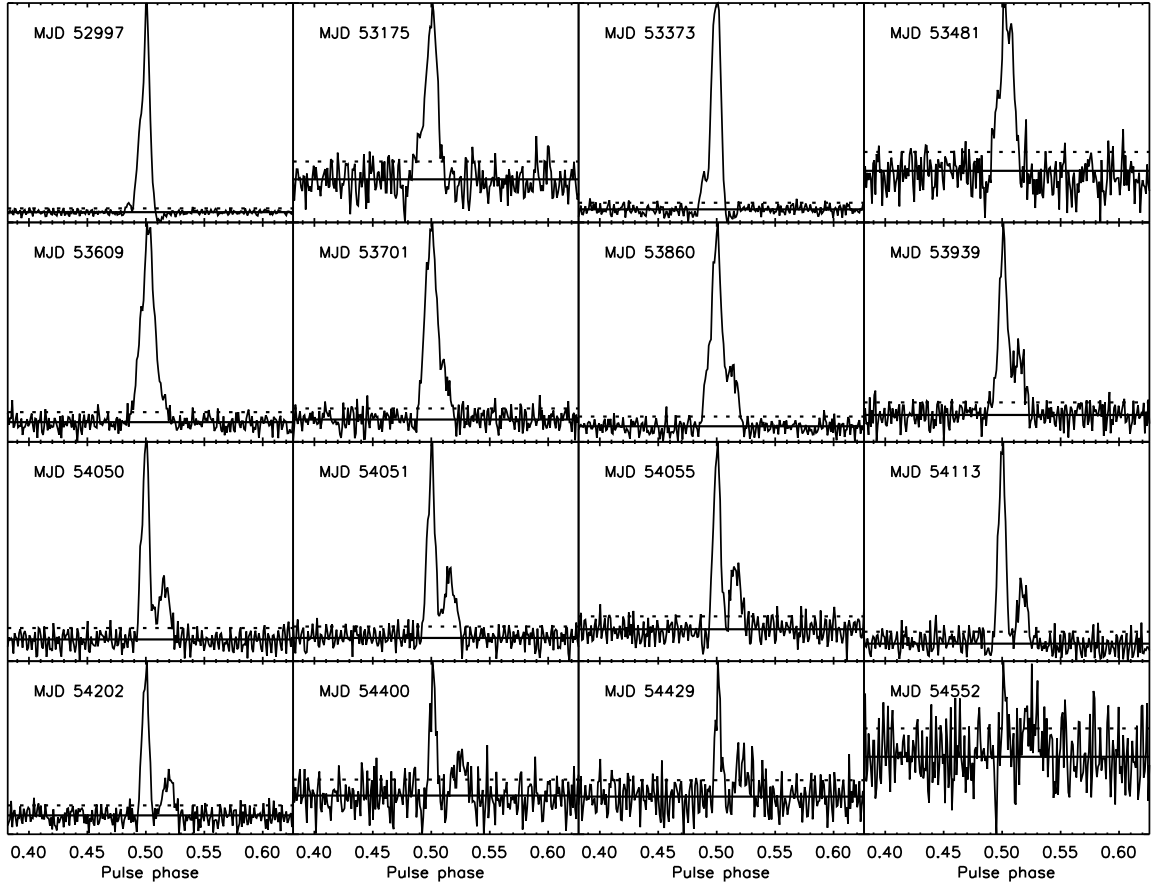}
\caption{
Mean pulse profiles for BP1 on 16 different days including the very first observation which was made on 2003 December 24 (MJD 52997). All data have been observed at a frequency of 820~MHz. There are 1024 bins across the entire pulse profile, resulting an effective time resolution of $0.003$~s. Since predictions of absolute pulse phase are not available for these observations, we aligned the maximum peak to the pulse phase of 0.5 at each epoch. The horizontal and dotted lines show the baseline, or off-pulse mean, of the profile and the standard deviation of the off-peak region, respectively. The signal-to-noise ratio (S/N) of pulse profiles has decreased significantly from 284 (on MJD 52997) to 11 (on MJD 54552).
\label{profiles}}
\end{figure*}

The mean pulse profiles for BP1 are shown in Figure~\ref{profiles}. We aligned the peak of the profiles to the pulse phase of 0.5 at each epoch. Note that these pulse profiles have a better resolution than those in \citet{pmk+10}. The second peak of the pulse profile can be hidden with the low time resolution. Therefore, we use 1024 bins across the full pulse phase, resulting in an effective time resolution of $0.003$~s in this analysis compared to $0.01$~s in \citet{pmk+10}. For example, the second peak of the pulse profile of MJD $53860$ around pulse phase $0.52$ in Figure~\ref{profiles} cannot be clearly seen in Figure~1 of \citet{pmk+10} on the same day with low time resolution. We use the pulse profiles from 23 days in section~\ref{remodel}, including 16 epochs in Figure~\ref{profiles}, in order to derive the beam shape. We include some low signal-to-noise data (e.g. MJD~53481) in this particular analysis because the second peak became apparent around those days. Because these profiles appeared as single-peaked, we ignored them in the geometrical modeling of the previous lower time resolution study.
As in \citet{pmk+10}, we fit one and two Gaussians for each single and double-peaked pulse profile, respectively, and then calculate profile widths at different intensity levels. We use these data in section~\ref{remodel} to determine the geometry of the pulsar and the beam.

\section{Boundary model}
\label{bmodel}

Due to the distortion of the magnetosphere, the properties of pulsar B are different from those of normal isolated pulsars. In isolated pulsars, we can determine the size of the magnetosphere by modeling the open and closed field lines, given the size of their light cylinder. However, as mentioned earlier, the magnetosphere of B is located deep inside the light cylinder and the structure of the open and closed field lines is more complicated due to the distortion from A's wind.

\begin{figure*}
\epsscale{1.50}
\plotone{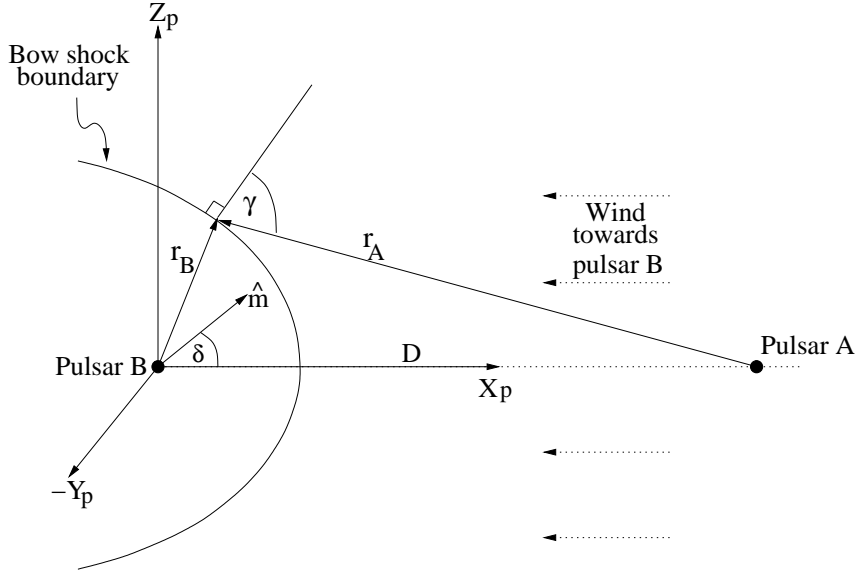}
\caption{
Geometry of the wind-magnetosphere interaction model. Pulsar A is located at a distance D ($\sim 9\times 10^{10}$ cm) away from  B along the $x_p$-axis. The wind of  A comes along the $-x_p$ direction and is shocked near  B. The physical interface has the shape of a bow shock and this is the boundary of the magnetosphere of B. ${\vec {r_B}}$ is the position vector at any point on the interface with respect to B. ${\vec {r_A}}$ is the position vector of this point with respect to  A and $\gamma$ is the angle between the normal to the boundary at this point and ${\vec {r_A}}$. $\delta$ is the angle between the magnetic moment axis $\hat{m}$ and the line connecting the two pulsars.
\label{boundary}}
\end{figure*}

In the first step, we approximate the structure of the magnetosphere as a rotating vacuum dipole. Then we apply a simple model for the wind-magnetosphere interaction, as in \citet{lyu04}. The wind of A creates a dynamic pressure on the magnetosphere of  B. The magnetosphere of  B creates a magnetic pressure which opposes the wind pressure of A. At some point, these two pressures equal each other; this interface is likely the boundary of the magnetosphere of  B. This boundary can be used to calculate the last open and closed field lines. We derive an expression for this boundary by equating the two pressures,

\begin{equation}
\label{pressure}
{\overrightarrow B}^{2}({\overrightarrow {r_B}})/(8\pi) = L_{A}\cos^2(\gamma)/4\pi c{\overrightarrow {r_A}}^2,
\end{equation}

\noindent
where ${\overrightarrow {r_B}}$ is the distance vector of the boundary with respect to pulsar B, ${\overrightarrow B}({\overrightarrow {r_B}})$ is the magnetic field of pulsar B at ${\overrightarrow r}_{B}$, $L_{A}$ is the spin-down luminosity of A, ${\overrightarrow {r_A}}$ is the distance vector of the boundary with respect to A, and $\gamma$ is the angle between the normal to the boundary and ${\overrightarrow {r_A}}$ (see Figure~\ref{boundary}). The relative pressures lead to a boundary much closer to B. For that reason we simplify the problem by setting the distance of the boundary with respect to A equal to the distance between the two pulsars and $\gamma$ equal to the angle between the normal to the boundary and the line connecting two pulsars. By assuming a magnetic dipole at the center of the coordinate system, we can write the magnetic field strength of the neutron star as ${\overrightarrow B}({\overrightarrow {r_B}}) = (3{\hat{r}_B}({\overrightarrow m} \cdot \hat{r}_B) - {\overrightarrow m})/{r_B}^3$, where ${\overrightarrow m} = m(\cos\delta\cos\Omega t,\cos\delta\sin\Omega t, \sin\delta)$ is the magnetic moment, $\Omega$ is the rotational angular frequency, and $\delta$ is the angle between the magnetic moment and the line connecting the two pulsars (see Figure~\ref{boundary}).

In isolated pulsars, we believe that the spin-down is caused by the power carried out along the magnetic field lines which are open with respect to the light cylinder \citep{cs06}. These open magnetic field lines start from the polar cap region of the neutron star surface and their shape changes with respect to the magnetic inclination. \citet{spi06} proposed a realistic form of the spin-down luminosity of an isolated pulsar as a function of the magnetic inclination angle. However, the open field line structure of pulsar B is somewhat different than for an isolated pulsar and, therefore, we need to define the magnetic fields lines with respect to the bow shock boundary. Therefore, we modified the spin-down luminosity equation that is given in \citet{spi06} by including the area of the polar cap region. This can be written as

\begin{equation}
\label{lumB}
L_{B} = \Omega^2 B_0^2 S^2 (1+\sin^2\alpha)/4\pi^2 c
\end{equation}

\noindent
where $\Omega$ is the rotational angular frequency, $B_0$ is the polar magnetic field of pulsar B, $S$ is the area of the polar cap region, and $\alpha$ is the magnetic inclination. As we mentioned earlier, with previous geometry models \citep{bkk+08,pmk+10}, $\alpha$ is taken to be $\sim 70\degr$. Note that the original version of the equation \citep[as in][]{spi06} can be obtained by taking $S$ to be  the area of the polar cap of a dipole as defined by the open magnetic field lines with respect to the light cylinder.

In order to determine the boundary, we solve equation (\ref{pressure}) numerically and then use equation (\ref{lumB}) to determine the value of the magnetic field. To do so, we simplified the problem to a 2D form in which  the bow shock is represented by an equation involving $x_p$ and $z_p$ and lies on that plane. The equation of the bow shock then has the form  $G(x_p,z_p) = f(z_p)-x_p$, which must be solved in order to determine the shape of the bow shock. Let the radial vector ${\vec {r_B}} = f(z_p){\hat{x}_p} + z{\hat{z}_p}$ and the vector normal to the boundary ${\vec n} = {\vec {\nabla}} \cdot G(x_p,z_p)$. Then the dot product of these two gives the angle $\cos^2(\gamma) = 1/(1+(df/dz_p)^2)$ and this can be substituted in equation (\ref{pressure}).
According to our 2D form, we can write the magnetic moment ${\vec m} = m\cos\delta{\hat{x_p}} + m\sin\delta{\hat{z_p}}$ and then derive the magnetic field of B at distance ${\vec {r_B}}$, ${\vec B}({\vec {r_B}})$, as a function of $m$, $f(z_p)$, $z$ and $\delta$. Then  equation (\ref{pressure}) reduces to a first order differential equation of $f(z_p)$ and the solution determines the shape of the bow shock. First we assume an initial value for the magnetic moment, $m$, of  B with a possible magnetic orientation, $\delta$, and solve the problem to determine the shape of the bow shock. Then we find which are the last closed field lines defined with respect to the bow shock and finally the shape and the area of the polar cap.

We repeat the procedure for 16 values of the angle $\delta$ between the magnetic moment and the line connecting the two pulsars, evenly spaced between 0 and $\pi/2$, and we find the area of the polar cap for each of those orientations. In the estimation of the average area of the polar cap, we have weighted appropriately the fact that some values of $\delta$ occur more frequently than others during an orbital period.
Using the spin-down luminosity  given in equation (\ref{lumB}) with the timing-derived $L_B$, we find a new value for the magnetic field. We repeat this process  with this new magnetic field until the value of the magnetic field converges. This happens after five to ten iterations for an initial guess of the magnetic field within a couple of orders of magnitude away from the convergence value. In order to represent the 3D version of the bow shock, we assume that it is axially symmetric around $x_p$.

According to the best solution, the magnetic field of B is constrained to be $B_{B} = 6.4\times10^{11}$ G, which is about a factor of two lower than the timing-derived value $1.2\times10^{12}$~G assuming a vacuum dipole with a magnetic inclination of $90\degr$ \citep{lbk+04}. This new estimate is more realistic as it accounts for the boundary of the magnetosphere as the bow shock and a realistic magnetic inclination. The stand-off distance is constrained to be either $3.8\times10^{9}$~cm or $4.5\times10^{9}$~cm for the cases when the magnetic axis is either normal or parallel to the line connecting the two pulsars, respectively.
Thus, the size of the boundary depends on the orientation of the magnetic axis of pulsar B. Moreover, the shape of the bow shock depends on the orientation of the magnetic axis.
The stand-off distance corresponds approximately to 1/3 of the light cylinder, thus for these distances the relativistic modifications are minimal and do not change the value of the stand-off distance by more than a few percent. For that reason we have chosen to calculate it using a vacuum dipole model rather than a more complicated geometry that takes into account relativistic effects as in \citet{deu55}.
For simplicity, we assume the boundary is axially symmetric around the vector connecting the two pulsars. A maximum deviation of roughly 20\% of the actual shape from the symmetric case occurs when the angle $\delta$ is $90\degr$.
Therefore, the shape of the boundary is sensitive to an angle of $\delta$, having a range of $[0\degr, 90\degr]$. If the angle $\delta$ is greater than $90\degr$, then the boundary considered the effective $\delta$ of $180\degr-\delta$. For example, if $\delta$ is $100\degr$, then the effective $\delta$ for the boundary shape is $80\degr$.
Therefore, with the assumption that the magnetic axis is nearly aligned with the line-of-sight at the radio emission detection, the effective angle $\delta$ is small ($\sim 5\degr-35\degr$) in the orbital longitude region of BP2. Thus, the deviation of the boundary model from the actual geometry of the boundary is small and the assumption of a symmetric geometry is reasonable. However, the deviation in the orbital phase region of BP1 is significant due to large effective angles of $\delta$ ($\sim 35\degr-85\degr$).

To derive an expression for the physical shape of the boundary, we examine different shapes which fit our results. As a preliminary fit, a parabola is a good guess, but a fourth-order polynomial describes the boundary better, yielding the  minimum chi-squared value when we fit to our results. The coefficients of this polynomial describe the variation and are functions of the angle $\delta$. The best-fit polynomial is

\begin{equation}
x_{p} = a(\delta) + b(\delta)(y_{p}^2 + z_{p}^2) + c(\delta)(y_{p}^2 + z_{p}^2)^2,
\end{equation}

\noindent
with a B-centered coordinate system in which the $x_{p}$ axis is towards A, the $z_{p}$ axis is normal to the orbital plane, and the $y_{p}$ axis  completes the right-handed coordinate system (see Figure~\ref{boundary}). The three axes have units in centimeters. The coefficients $a(\delta)$, $b(\delta)$, and $c(\delta)$ are

\begin{eqnarray}
a(\delta) &=& \left( \frac{0.83 - 0.01\delta - 0.06\delta^2 - 0.05\delta^3 + 0.03\delta^4}{1.83\times10^{-10}} \right) \nonumber\\
b(\delta) &=& \left(\frac{-0.46 + 0.04\delta - 1.43\delta^2 + 1.96\delta^3 - 0.64\delta^4}{5.45\times10^9}\right) \\
c(\delta) &=& \left(\frac{-0.48 - 0.03\delta + 2.15\delta^2 - 2.47\delta^3 + 0.74\delta^4}{1.62\times10^{29}} \right) \nonumber
\end{eqnarray}

\noindent
where the angle $\delta$ is in radians and having a range of $[0,\pi/2]$. Since they are functions of $\delta$, the boundary changes slightly with spin and orbital motions, as well as over time due to precession.

This boundary model is valid only up to $5\times 10^{9}$~cm, or $\sim 40\%$ of the light cylinder radius, from B. Beyond this limit, the physical assumption of the dynamical pressure is incorrect because the wind pressure on the magnetic field should be zero when it is parallel to the boundary at large distances. Also we have assumed an undistorted magnetic field of B in the model and the distortions at large distances will be significant.

In summary, we determined the boundary of pulsar B by assuming equilibrium between the dynamical pressure of the wind of A and the magnetic pressure of the field of B. The physical shape of the boundary is a bow shock and mathematically we can represent it as a fourth-order polynomial. Moreover, this shape depends on the orientation of the magnetic moment axis with respect to pulsar A. Thus, the coefficients of the best fit polynomial depend on this orientation.

\subsection{Tracing the dipole field lines}
\label{field_lines}

Due to the wind interaction with the magnetosphere of B, it is complicated to understand the structure of the open and closed field lines. To determine the polar cap region that is required for the emission height estimation, we calculate the last closed field lines by tracing them referring to the derived boundary model.

As is standard, we treat the magnetosphere of B as a magnetic dipole. For an isolated pulsar, the last closed magnetic field lines are defined as those that just touch the light cylinder and the ones interior to the last closed field lines are considered open field lines. In our case the boundary is not the light cylinder but the bow shock, with the last closed field lines defined as those that just touch this bow shock.  The polar cap region is defined by these particular field lines and the shape of it can be determined by the locations where these field lines cross the neutron star surface. Defining the polar cap is important since we think that the radio emission is produced above this region.

Unlike those of isolated pulsars, the magnetosphere of B is not symmetric around the magnetic axis due to the shape of the  boundary. This can be clearly seen by tracing the last closed field lines.
In order to trace the field lines, we use the dipole field line equation in polar coordinates

\begin{eqnarray}
\label{dipole}
r &=& r_{0}\sin^{2}(\lambda) \nonumber\\
\phi &=& \phi_{0}
\end{eqnarray}

\noindent
where $r$ is the radial distance to a given point along the field line and $r_{0}$ is the field-line constant, or equatorial distance of the field line from the magnetic axis. The angle $\lambda$ is the colatitude of a given point along the field line and $\phi_{0}$ is the azimuth angle, or the longitude of the given field line. Then the Cartesian components of a particular field line are written as

\begin{eqnarray}
\label{corotate}
x &=& r\sin(\lambda)\cos(\phi) \nonumber\\
y &=& r\sin(\lambda)\sin(\phi) \\
z &=& r\cos(\lambda) \nonumber
\end{eqnarray}

\noindent
where the $z$-axis of the coordinate system is aligned with the magnetic moment axis and the other two axes are co-rotating with the neutron star.

\begin{figure*}
\epsscale{1.8}
\plotone{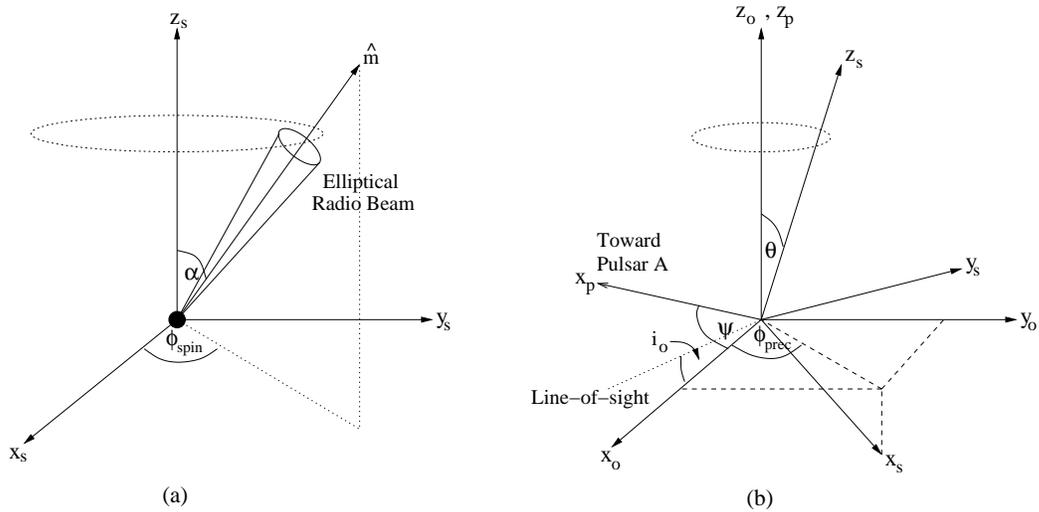}
\caption{
Cartesian coordinate systems that transform magnetic field lines from the co-rotating frame of the neutron star to the orbit-fixed frame. (a) The spin phase $\phi_{spin}$ and the misalignment of the magnetic axis $\alpha$ are defined in the frame $(x_{s},y_{s},z_{s})$. The spin phase is measured from the $x_s$-axis. Therefore $\phi_{spin}=0$ is defined when the magnetic axis is in $x_{s}$--$z_{s}$ plane. (b) The colatitude of the spin $\theta$ and the precessional phase $\phi_{prec}$ is defined in the frame $(x_{o},y_{o},z_{o})$. In this frame, the $x_{o}$-axis is in the plane of the $z_{o}$-axis and our line-of-sight (LOS). The spin precession $\phi_{prec}$ is measured from the $x_o$-axis. The frame of the bow shock boundary ($x_p, y_p, z_p$) is then placed in this orbit-fixed frame with $z_p \parallel z_o$ and rotate with an angle of $\psi$ defined from the $x_o$-axis. Then the orbital phase is defined $\phi_{orb} = \psi + 90\degr$ as it measures from the ascending node. $i_o=90\degr-i$, where $i$ is the orbital inclination.
\label{coord}}
\end{figure*}

To include the misalignment of the magnetic axis and also account for the spin phase, we transform equation (\ref{corotate}) to another frame where the $z$-axis is aligned with the spin axis. In this frame (see Figure~\ref{coord}(a)), the Cartesian components are

\begin{eqnarray}
x_{s} &=& x\cos\alpha\cos\phi_{spin} - y\sin\phi_{spin} + \nonumber\\
	&& z\sin\alpha\cos\phi_{spin} \nonumber\\
y_{s} &=& x\cos\alpha\sin\phi_{spin} + y\cos\phi_{spin} + \nonumber\\
	&& z\sin\alpha\sin\phi_{spin} \\
z_{s} &=& z\cos\alpha - x\sin\alpha \nonumber
\end{eqnarray}

\noindent
where $\alpha$ is the angle between the magnetic axis and the spin axis and $\phi_{spin}$ is the spin phase. We measure the spin phase from the $x_{s}$ axis, which means that it is zero when the magnetic axis is in the $x_{s}$--$z_{s}$ plane.

The spin axis is also associated with the colatitude angle and the spin precessional phase. Geodetic spin precession changes the orientation of the spin axis with respect to our line-of-sight. In order to include the colatitude of the spin axis and effects of spin precession, we transform a particular field line to another frame which is fixed with our line-of-sight. We choose the frame with $z$-axis parallel to the orbital angular momentum axis and $x$-axis in the plane of the $z$-axis and the line-of-sight (see Figure~\ref{coord}(b)). The Cartesian components in this frame are

\begin{eqnarray}
x_{o} &=& x_{s}\cos\theta\cos\phi_{prec}(t) - y_{s}\sin\phi_{prec}(t) + \nonumber\\
	&& z_{s}\sin\theta\cos\phi_{prec}(t) \nonumber\\
y_{o} &=& x_{s}\cos\theta\sin\phi_{prec}(t) + y_{s}\cos\phi_{prec}(t) + \nonumber\\
	&& z_{s}\sin\theta\sin\phi_{prec}(t) \\
z_{o} &=& z_{s}\cos\theta - x_{s}\sin\theta, \nonumber
\end{eqnarray}

\noindent
where $\theta$ is the angle between the spin axis and the orbital angular momentum axis and $\phi_{prec}(t)$ is the spin precession phase measured from the $x$-axis (i.e., $\phi_{prec}(t) = 0$ when the spin axis is in the plane of $x_{o}$--$y_{o}$). The spin precession phase is given by

\begin{equation}
\label{prec}
\phi_{prec}(t) = \Omega_{prec}(t - T_0)
\end{equation}

\noindent
where $\Omega_{prec}$ is the spin precession rate of B, which is 5.061(2)$^{\circ}$~yr$^{-1}$, and $T_0$ is the time when the $\phi_{prec}(t)$ is zero, defined as the time when the spin axis is in the $x_{o}$--$y_{o}$ plane.
By using the above set of equations, we can transform dipole field lines from the co-rotating frame of the neutron star to the orbit-fixed frame $(x_{o},y_{o},z_{o})$ where the $x_{o}$--$y_{o}$ plane is in the orbital plane and $x_{o}\sin(i)$ is pointing towards the observer, where $i$ is the orbital inclination.

\begin{figure*}
\epsscale{2.0}
\plotone{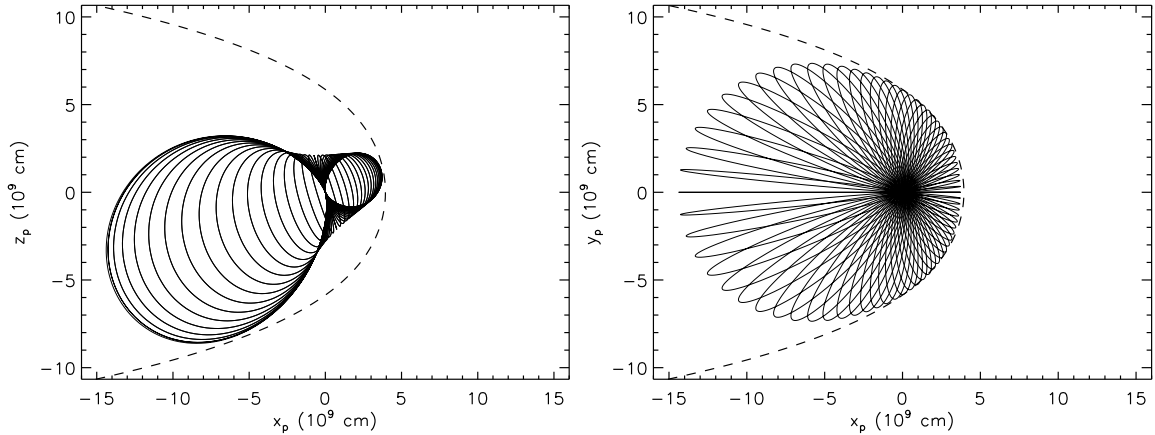}
\caption{
Confined rotating vacuum dipole in the boundary model on MJD 54050 (2006 November 11) $-$ a view in the $x_p$--$z_p$ plane (left) and a view in the $x_p$--$y_p$ plane (right). Pulsar B is located at the center of the coordinate system and the wind of  A comes towards the $-x_p$ direction. The dashed line shows the derived bow shock from the wind-magnetosphere interaction model and this models the open and closed field lines of the magnetosphere. The solid lines are the last closed field lines with respect to this boundary. The field lines that  have higher latitude than these shown closed field lines are considered open field lines. The scale is in units of $10^{9}$ cm. Here, $\alpha=61\degr$ and $\theta=138.5\degr$; these are our best-fit geometrical parameters from section~\ref{remodel}. This view corresponds to the orbital phase of 223$\degr$. Note, for clarity of plots, we take the spin phase as zero, so that the north pole of the pulsar is pointing below the $z_p=0$ plane. However, the shape of the magnetosphere changes with the spin and orbital motion and over time due to precession.
\label{parabola}}
\end{figure*}

In order to place the polynomial boundary in the $(x_{o},y_{o},z_{o})$ frame, we need to account for the orbital motion of B. Due to this motion, the orientation of the boundary changes with respect to our line of sight, because the location of A changes with respect to B. This relative motion changes the shape of the magnetosphere of B with respect to the line-of-sight and then the shape of the polar cap region. This results a variation in the emission height estimate across the orbit (more details are given in section~\ref{aceh}). We place the boundary model $(x_{p},y_{p},z_{p})$ in this orbit-fixed coordinate frame with $z_{o}\parallel z_{p}$ and then rotate it corresponding to the orbital phase $\phi_{orb}$ with $(x_p \cos\phi_{orb} - y_p\sin\phi_{orb}, x_p\sin\phi_{orb} + y_p\sin\phi_{orb},z_p)$, where $\phi_{orb} = \psi + 90\degr$ (see Figure~\ref{coord}(b)). Here $\phi_{orb}$ is measured from the ascending node and $\psi$ is the rotation angle between $x_o$ and $x_p$ axes. Note that this $90\degr$ angle is included to convert the rotation angle $\psi$ to orbital phase $\phi_{orb}$ as measured from the ascending node. Then we trace the last closed field lines according to the orientation of the boundary for the given orbital phase. For example, Figure~\ref{parabola} shows the confined magnetosphere in the boundary model with the last closed field lines on MJD 54050 (2006 November 11). Here, we use our best-fit geometry of the pulsar from the beaming model that is described in section~\ref{remodel}. At this particular epoch, the spin precession phase is $\phi_{prec} = 46\degr$ and we use the orbital phase $\phi_{orb}=223\degr$ and the spin phase $\phi_{spin}=0$ in the figure. Note that, at this spin and the orbital phases, the spin axis of the pulsar is in the $x_p$--$z_p$ plane.

The spin of the magnetic axis also results in a change in the shape of the magnetosphere due to the misalignment of the magnetic moment. However, the most important orientation of the magnetic axis is when it reaches the closest approach to our line-of-sight (i.e. where we detect the emission). In order to measure the point of this closest approach, or the impact parameter $\beta(t)$, we use the equations

\begin{eqnarray}
\label{impact}
&& \cos\zeta(t) = \sin\theta\cos\phi_{prec}(t)\sin i + \cos\theta\cos i \nonumber \\
&& \beta(t) = \zeta(t) - \alpha.
\end{eqnarray}

\noindent
Here, $\zeta(t)$ is the angle between the spin axis and our line-of-sight at a given time and the other angles have the usual meaning. We calculate the spin phase which gives this particular closest approach, so that we can estimate the emission height only at this particular spin phase.

Now we can transform field lines from the co-rotating frame of the neutron star to the orbit-fixed frame. By using the boundary model, we can trace the last closed field lines, which determine the shape of the magnetosphere. Due to different orientations of the magnetic axis with spin, the shape of the magnetosphere with respect to our line-of-sight changes, but we are interested only in the spin phase which gives the closest approach to us. Nevertheless, orbital motion and spin precession change the shape and we need to account for these in emission height calculation.

\section{Re-analysis of the beam shape}
\label{remodel}

For the emission height calculation, we require the angular radius of the emission beam. \citet{pmk+10} claimed the beam shape of B to be elliptical by modifying the \citet{cw08} geometrical framework.
The model in \citet{pmk+10} used the 2D geometry of the beam after projecting it to a plane which is perpendicular to the spin axis, resulting in a projected angular radius. We improve this model by using a more realistic 3D model in this paper in order to determine the actual angular radius.

\begin{figure*}
\epsscale{1.0}
\plotone{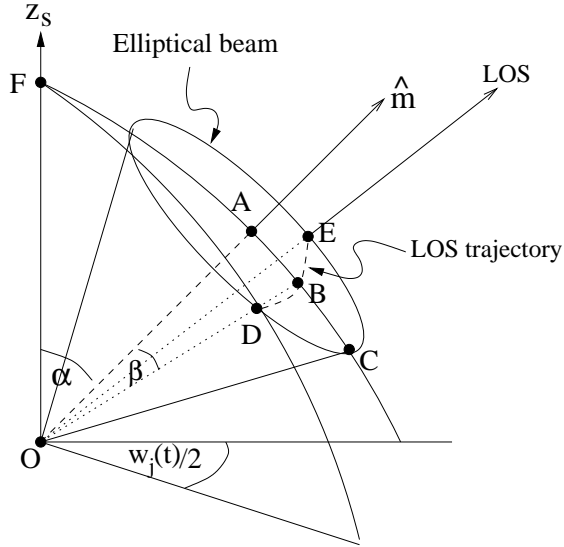}
\caption{
Elliptical emission cone in the frame $(x_s,y_s,z_s)$. The magnetic axis of the beam represents with $\hat{m}$ and it is misaligned with the spin axis of an angle $\alpha$. The trajectory of the line-of-sight across the beam due to rotation is denoted with $DBE$. The angle $B\hat{O}D$ is the longitudinal angular radius $\rho_{l,j}(t)$ of the beam for a given intensity level at a given time. The angle $A\hat{O}C$ is the angular radius across the semi-major axis of the beam $\rho_{a,j}$ for a given intensity level, which is time independent and fixed for the beam. The angle $A\hat{O}D$ is the effective angular radius of the beam $\rho_{e,j}(t)$ for a given intensity level.
\label{beam}}
\end{figure*}

The angular radius of a circular emission beam has been calculated using pulse profile widths and an assumed emission geometry \citep{ggr84}. Since  \citet{pmk+10} showed that the shape of B's beam is not circular, we attempt to derive an approximate equation for an elliptical beam in 3D and then follow the same analysis of \citet{pmk+10} in order to determine the geometry.
First, we construct the beam with a set of coaxial hollow cones representing different intensity levels in a way such that the outermost one represents the lowest intensity level and then the intensity increases gradually inwards toward the maximum and then decreases until reaching the center of the beam. All these cones have cross sections with a constant ellipticity of $a_j/b_j$, where $a_j$ and $b_j$ are semi-major and minor axes, respectively, of each hollow-cone beam. Then, for any given longitudinal angular radius $\rho_{l,j}(t)$ (see Figure~\ref{beam}), magnetic misalignment angle $\alpha$, and impact parameter $\beta(t)$, the pulse profile width $w_{j}(t)$ can be derived from spherical trigonometry (i.e. from spherical triangle FBD in  Figure~\ref{beam}) as follows

\begin{equation}
\label{width}
w_{j}(t) = 2 \arccos \left( \frac{\cos(\rho_{l,j}(t)) - \cos^2(\alpha+\beta(t))}{\sin^2(\alpha+\beta(t))} \right),
\end{equation}

\noindent
where subscript $j$ specifies different intensity levels of the pulse profile.
Note that $\rho_{l,j}(t)$ is time dependent because the region where our line-of-sight cuts the beam is changing with time due to precession.
In order to relate $\rho_{l,j}(t)$ with the elliptical beam shape, we derive an equation with the assumption that the cross-section of the beam is small enough to use 1D trigonometry. Then the longitudinal angular radius $\rho_{l,j}(t)$ can be given as a function of $\beta(t)$,

\begin{equation}
\label{radius}
\rho_{l,j}(t) = \frac{1}{\chi}\sqrt{\sin^{2}(\rho_{a,j}) - \cos^{2}(\rho_{a,j})\tan^{2}(\beta(t))},
\end{equation}

\noindent
where $\rho_{a,j}$ is the angular radius across the semi-major axis of the beam (see Figure~\ref{beam}) for a given intensity level and $\chi=a_j/b_j$, which is a constant for all different intensity cones.
This expression shows that the minimum $\rho_{l,j}(t)$ of zero occurs when the line-of-sight just encounters the beam (i.e. $\beta(t) = \rho_{a,j}$), resulting in $w_{j}(t) =0$. The maximum $\rho_{l,j}(t)$ occurs when the line-of-sight crosses the center of the beam (i.e. $\beta(t)=0$), which leads to the maximum $w_{j}(t)$. Therefore, by combining equation~(\ref{width}) and (\ref{radius}) for a given $\alpha$, $\beta(t)$, and $\rho_{a,j}$, we can calculate the pulse profile width $w_j(t)$ for any given intensity level.

\begin{figure*}
\epsscale{1.50}
\plotone{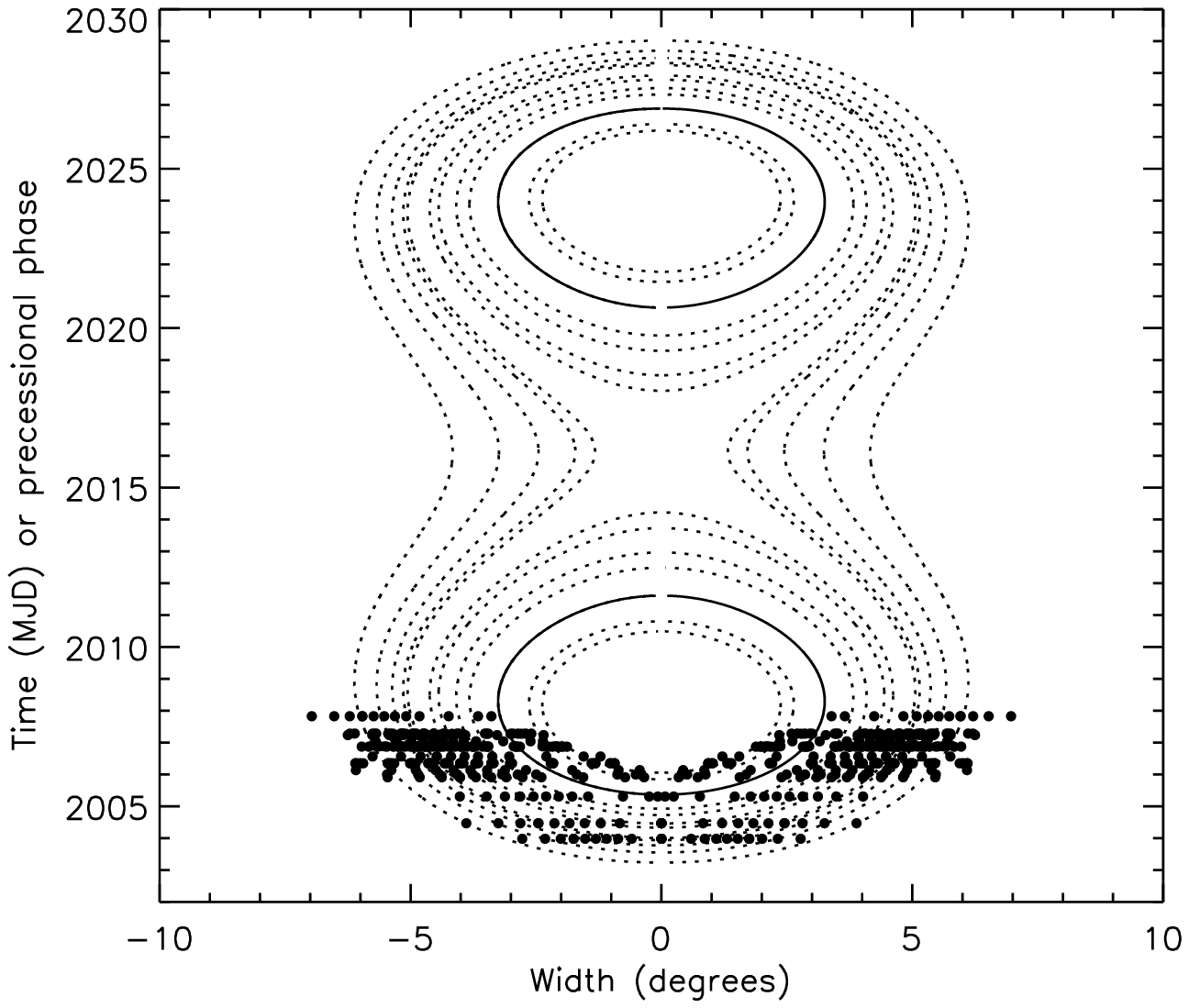}
\caption{
2D pulse profile of the orbital longitude region $185\degr-235\degr$ (BP1), assuming the elliptical hollow-cone beam. The best-fit geometrical parameters are $\alpha=61.0\degr_{-2.4^{\circ}}^{+7.9^{\circ}}$, $\theta=138.5\degr_{-4.4^{\circ}}^{+5.3^{\circ}}$ and $\chi=2.6_{-0.6}^{+0.4}$ (errors are 1$\sigma$). The corresponding $T_0$ is MJD 57399 (2006 November 11), which is the time where the spin axis of the pulsar is in the plane of our line-of-sight and the orbital angular momentum axis. Note that these $\alpha$ and $\theta$ values are consistent with previous results \citep{pmk+10,bkk+08}. The dots are the widths at equal intensity levels of the observed pulse profile. Each horizontal row of dots represents an observation at a given epoch. Equal-intensity contours are the elliptical beam model-predicted pulse profile widths at different intensities. The intensity increases from the inner dashed line  outwards until the first solid line, which is the intensity of the peak, and then decreases outwards again. The intensity levels are, from the inner dashed line, 80\%, 90\%, 100\%, 90\%, 80\%, 70\%, 60\%, 50\%, 40\%, 30\%, 20\% and 10\%. The vertical axis is calibrated in years and can also be considered the spin precession phase.
Note that, this figure shows the emission from the full elliptical beam. If the beam is partially-filled according to 2008 radio disappearance, then the model predicts no radio pulse profiles from 2008 to 2024.
\label{model}}
\end{figure*}

In order to determine the geometry of B, we fit the model-predicted pulse profile widths to observed pulse profile widths of BP1 at different intensity levels using the same likelihood analysis that was described in \citet{pmk+10}. The fit was done by searching the entire parameter space of $\alpha$, $\theta$, $\chi$, and $T_0$. For each combination of these parameters, we vary $\rho_{a,j}$ from $0\degr$ to $30\degr$ freely until we reach the best solution. Then we use a maximum likelihood analysis to determine the best-fit geometrical parameters. The best-fit model for BP1 is shown in Figure~\ref{model}. The estimated geometrical parameters $\alpha =61.0\degr_{-2.4^{\circ}}^{+7.9^{\circ}}$ and $\theta=138.5\degr_{-4.4^{\circ}}^{+5.3^{\circ}}$ are consistent with those in \citet{pmk+10} and \citet{bkk+08} within the 2-$\sigma$ errors. The ratio $\chi$ is constrained to be $2.6_{-0.6}^{+0.4}$, lower than the estimate of the previous paper. Our new estimate is more believable because it has been derived from a full 3D viewing model. In addition to these parameters, we derive $T_0$ to be MJD $57399_{-25}^{+4}$ (2016 January 12), which results in a precessional phase of $46\degr$ at an epoch of MJD 54050 (2006 November 11). This estimate is consistent with the value predicted by \citet{bkk+08} at the same epoch. However, this is a somewhat arbitrary parameter that can be chosen from our best-fit model. Note that the best-fit $T_0$ in \citet{pmk+10} is about MJD 33360 (1950 March 20), which results in a precessional phase of $73\degr$ at an epoch of MJD 54050. These two best-fit $\phi_{prec}$ result in a shift of the hour-glass 2D pulse profile shape along the precessional phase or time axis \citep[see Figure~\ref{model} of this analysis and Figure~17 of][]{pmk+10}.

The angular radius across the semi-major axis of the beam at the maximum intensity level $\rho_{a,100}$ and 10\% of the maximum $\rho_{a,10}$ are constrained to be $9.9\degr$ and $14.3\degr$, respectively.
In order to determine the effective angular radius of the beam (more details are given in section~\ref{aceh}), we use these angular radii with the derived beam geometry.

For
 emission height estimates for normal non-precessing pulsars, a beam shape is not essential because our line-of-sight always observes the same section of the emission beam. However, for precessing pulsars, we must consider a beam shape in order to determine the emission height due to observing different sections of the emission beam. Therefore, in this particular case, we use our derived elliptical beam shape with the best-fit geometrical parameters of B to estimate emission heights in section~\ref{aceh}. As we see in \citet{pmk+10}, the pulse profile evolution is somewhat similar in both bright phases. Therefore, we use the above best-fit beam parameters from BP1 in our emission height estimates for both bright phases.

\citet{pmk+10} reported that the radio emission of B disappeared in 2008 March, because the line-of-sight precessed away from the partially radio-filled elliptical beam.
According to the same argument that the beam is not entirely radio loud, we can explain the radio disappearance with our new elliptical beam geometry. With the partially filled beam configuration, we can predict the reappearance of the radio emission as our line-of-sight precess back to the radio loud region of the beam.
With the model described in \citet{pmk+10}, the reappearance is predicted to occur in around $2035$ with the same part of the beam or in around $2014$ if the beam has two symmetric radio-filled portions. However, our new model describes the reappearance is predicted to occur in around $2024$ with the same part of the beam. Therefore, Figure~\ref{model} changes with the partially-filled horse-shoe beam to non-detectable emission from 2008 to 2024.
If our new beam model is correct, then the beam should not have two symmetric radio-filled parts, because our line-of-sight crossed around the center of the beam when the disappearance occurred in 2008. Therefore, if there are two symmetric parts, then we would be able to detect radio emission at present day. The two different predictions for the reappearance from the two models mainly occur due to two different best-fit $T_0$ values. These two different $T_0$ values give two different solutions for the geometry of B, however, the best-fit values for parameters $\alpha$ and $\theta$ are the same for the two models with the 2-$\sigma$ uncertainty.

\section{Emission height calculation}
\label{aceh}

In order to estimate the radio emission heights of pulsar B, we use the previously defined boundary model, the field line tracing technique, and the modeled geometry of the beam. As we mentioned earlier, we assume that the radio emission is produced above the polar cap region and originates tangential to the local magnetic field lines.

First, we need to determine the boundary of the polar cap region, given by the last closed field lines. This can be done by tracing the field lines with the derived boundary model as described in section~\ref{field_lines}. We assume that the radio emission comes from above the entire polar cap region, so that the outer edge of the pulse profile (i.e. 10\% of the maximum intensity) corresponds to the region between the open and closed field lines, approximately equal to the last closed field line. Then we determine the emission height that originates from these last closed field lines.

In order to determine the tangent to a particular last closed field line at a given moment, we rewrite the coordinate transformations in section \ref{field_lines} in matrix form \citep[see, e.g.,][]{gan04}, so that it is easy to evaluate the equations relevant for our calculation. First, we write the dipole field equation in the co-rotating frame of the neutron star as

\begin{equation}
\overrightarrow{r}_{cor} = r_{0}(\sin^3\lambda\cos\phi, \sin^3\lambda\sin\phi, \sin^2\lambda\cos\lambda).
\end{equation}

\noindent
We then transform it to the orbit-fixed frame

\begin{equation}
\overrightarrow{r}_{orb} = A \cdot B \cdot \overrightarrow{r}_{cor}
\end{equation}

\noindent
where,

\begin{equation}
A =
\left( {\begin{array}{ccc}
 \cos\alpha\cos\phi_{spin} & -\sin\phi_{spin} & \sin\alpha\cos\phi_{spin}  \\
 \cos\alpha\sin\phi_{spin} & \cos\phi_{spin} & \sin\alpha\sin\phi_{spin}  \\
 -\sin\alpha & 0 & \cos\alpha \\
 \end{array} } \right)
\end{equation}

\noindent
and

\begin{equation}
B =
\left( {\begin{array}{ccc}
 \cos\theta\cos\phi_{prec} & -\sin\phi_{prec} & \sin\theta\cos\phi_{prec}  \\
 \cos\theta\sin\phi_{prec} & \cos\phi_{prec} & \sin\theta\sin\phi_{prec}  \\
 -\sin\theta & 0 & \cos\theta \\
 \end{array} } \right).
\end{equation}

\noindent
For the detection of radio emission, our line-of-sight must be parallel to the tangential vector of the given field line at a particular point. By locating this point on the field line, we can determine the height of the radio emission. To evaluate the tangent to the field line, we take $\overrightarrow{r}_{t} = \partial\overrightarrow{r}_{orb}/\partial\lambda$. Then the unit vector along the tangential direction ($\hat{r}_{t} = \overrightarrow{r}_{t}/|\overrightarrow{r}_{t}|$) can be written as,

\begin{equation}
\hat{r}_{t} = A \cdot B \cdot \sqrt{\frac{2}{5+3\cos(2\lambda)}} \left( {\begin{array}{c} 3\cos\phi\sin\lambda\cos\lambda \\ 3\sin\phi\sin\lambda\cos\lambda \\ 2\cos^2\lambda - \sin^2\lambda \\ \end{array} } \right).
\end{equation}

\noindent
The direction of the magnetic moment axis in the orbit-fixed frame can be written as

\begin{equation}
\hat{m} = A \cdot B \cdot \hat{z}_{o}.
\end{equation}

\noindent
Then we evaluate the angle between the direction of the magnetic moment axis and the vector tangential to the field line ($\tau$) at any time through the expression

\begin{equation}
\label{taueq}
\cos(\tau) = \hat{r}_{t} \cdot \hat{m} = \frac{1+3\cos(2\lambda)}{\sqrt{10+6\cos(2\lambda)}}.
\end{equation}

\noindent
This is the same as equation (8) in \citet{gan04}. At the point of detection of radio emission, we take the angle $\tau$ to be equal to the effective angular radius of the previously derived emission beam at the given time. According to our assumption that the outer edge of the pulse profile (10\% maximum) comes from the last closed field line, we take $\tau\approx \rho_{e,10}(t)$, where $\rho_{e,10}(t)$ is the effective angular radius of the beam (angle $A\hat{O}D$ of Figure~\ref{beam}) at the 10\% of the maximum intensity level corresponding to a particular impact parameter $\beta(t)$. We can derive an equation for $\rho_{e,10}(t)$ by using the spherical triangle $FAD$ of Figure~\ref{beam} as

\begin{eqnarray}
\cos(\rho_{e,10}(t)) = \cos(\alpha)\cos(\alpha+\beta(t)) + \nonumber \\
\sin(\alpha)\sin(\alpha+\beta(t))\cos(w_{10}(t)/2).
\end{eqnarray}

\noindent
Here, $w_{10}(t)$, the model-estimated pulse width, can be found through equation (\ref{width}) and (\ref{radius}) with the best-fit parameters $\alpha$, $\theta$, $\chi$, and $\rho_{a,10}$. The impact parameter $\beta(t)$ for the given time can be determined through equation (\ref{prec}) and (\ref{impact}) with the best-fit $T_0$.
This $w_{10}(t)$ is simply the 10\% pulse width of the 2D pulse profile given in Figure~\ref{model} at the given time. By simplifying equation (\ref{taueq}), we find an expression for $\lambda$, which is the colatitude of the emission point. This expression can be written

\begin{eqnarray}
\cos(2\lambda) = \frac{1}{3} [ \cos(\rho_{e,10}(t))\sqrt{8+\cos^2(\rho_{e,10}(t))} - \nonumber \\
\sin^2(\rho_{e,10}(t))].
\end{eqnarray}

\noindent
This is same as equation (9) in \citet{gan04}, so that the colatitude angle of the emission point in our complicated geometry is simplified to that of an isolated pulsar. Then the emission height of this point can be calculated by using the first equation of (\ref{dipole}). However, determining the field line constant, $r_{0}$, in this equation is difficult due to the bow shock boundary and its variation. \citet{kg97} assumed that this $r_{0}$ is the light cylinder radius since the isolated pulsars that they have studied have low magnetic inclinations. To determine $r_{0}$ for our particular case, we trace the last closed field line which is tangent to our line-of-sight at the closest approach of the magnetic moment with respect to the bow shock boundary. Then we use the dipole field equation to estimate the emission height.

\begin{figure*}
\epsscale{1.50}
\plotone{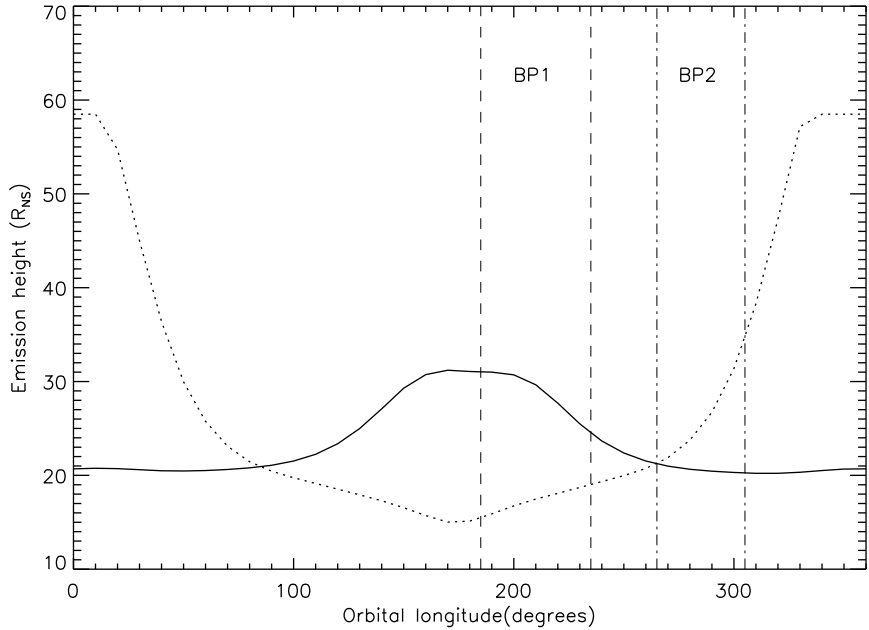}
\caption{
The model-estimated radio emission height across the orbit on MJD 54050 (2006 November 11). The solid line represents the height of the emission produced by the leading edge of the beam and the dotted line represents that for the trailing edge of the beam. In order to consistent with observations, we used our best-fit geometrical parameters of the beaming model, $\alpha=61\degr$ and $\theta=138.5\degr$. At this epoch, $\phi_{prec}$ is $46\degr$, the impact parameter $\beta$ is $3\degr$ and the corresponding spin phase is $126\degr$. Note that the difference between the corresponding heights for the leading and trailing components of the beam is more significant in some parts of the orbit . The orbital longitude regions for BP1 and BP2 are denoted with dashed and dot-dashed lines, respectively.
\label{orb}}
\end{figure*}

The orientation of the bow shock changes across the orbit with respect to our line-of-sight at the closest approach, so that the emission height changes with orbital longitude, because the last closed field lines are defined with respect to the bow shock. Moreover, when the central part of the radio beam crosses our line-of-sight, we will detect a double-peaked profile since the two edges, leading and trailing, of the beam cross our line-of-sight. Thus due to the different orientation of the magnetic moment axis at these two edges with respect to us, our line-of-sight is tangent to two different last closed field lines which have two different $r_{0}$ values. Thus the height of the emission produced by the leading and trailing edges of the beam are different. This is shown in Figure~\ref{orb}. For example, the emission heights produced by the leading components of the beam are constrained to be in a range of [$24\pm8$,$31\pm10$] and [$20\pm6$,$21\pm7$] in NS radii (10~km) on MJD 54050 (2006 November 11) for BP1 and BP2, respectively. The heights of the trailing edge of the beam in BP1 and BP2 are constrained to be in a range of [$15\pm5$,$19\pm6$] and [$21\pm7$,$38\pm12$] in NS radii, respectively. The errors of the height estimates are calculated from the 1-$\sigma$ uncertainties of the best-fit geometrical parameters from the beaming model. Thus, the uncertainty of the height estimate is in a range of [6,10] and [5,19] in NS radii for the leading and the trailing edge of the beam, respectively, across the orbit on this particular day.

\begin{figure*}
\epsscale{1.50}
\plotone{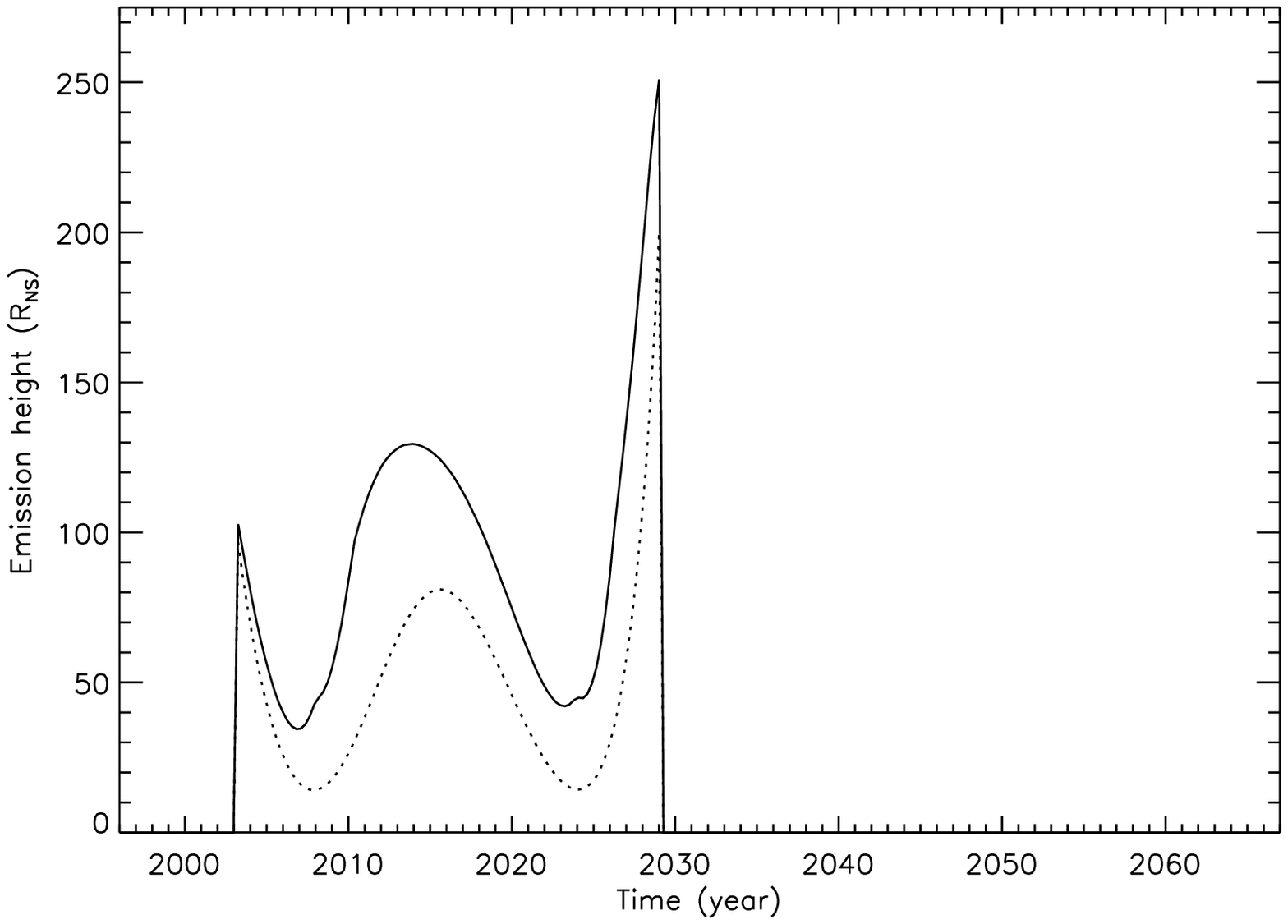}
\caption{
The model-estimated radio emission heights of the orbital longitude region of BP1 vs time across one precessional cycle. This predicted variation of emission height with time is due spin precession, making different line-of-sight cuts across the radio beam. The solid line represents the emission height from the leading edge of the beam and the dotted line represents that for the trailing edge. Here, the orbital phase is fixed at $200\degr$ (BP1), but the spin phase changes with time corresponds to the $\beta$ value. The time axis represents a full precession cycle, 71~yr.
Here, we have used the same geometrical parameters as in Figure~\ref{orb} and assumed a full elliptical beam, not a partially filled horse-shoe beam. If the beam is partially-filled then no radio emission is expected from 2008 to 2024. This is why the model still predicts radio emission at present-day MJDs. The emission height is zero ($\sim 2030-2067$) when the line-of-sight is out of the radio beam.
\label{mjd_BP1}}
\end{figure*}

\begin{figure*}
\epsscale{1.50}
\plotone{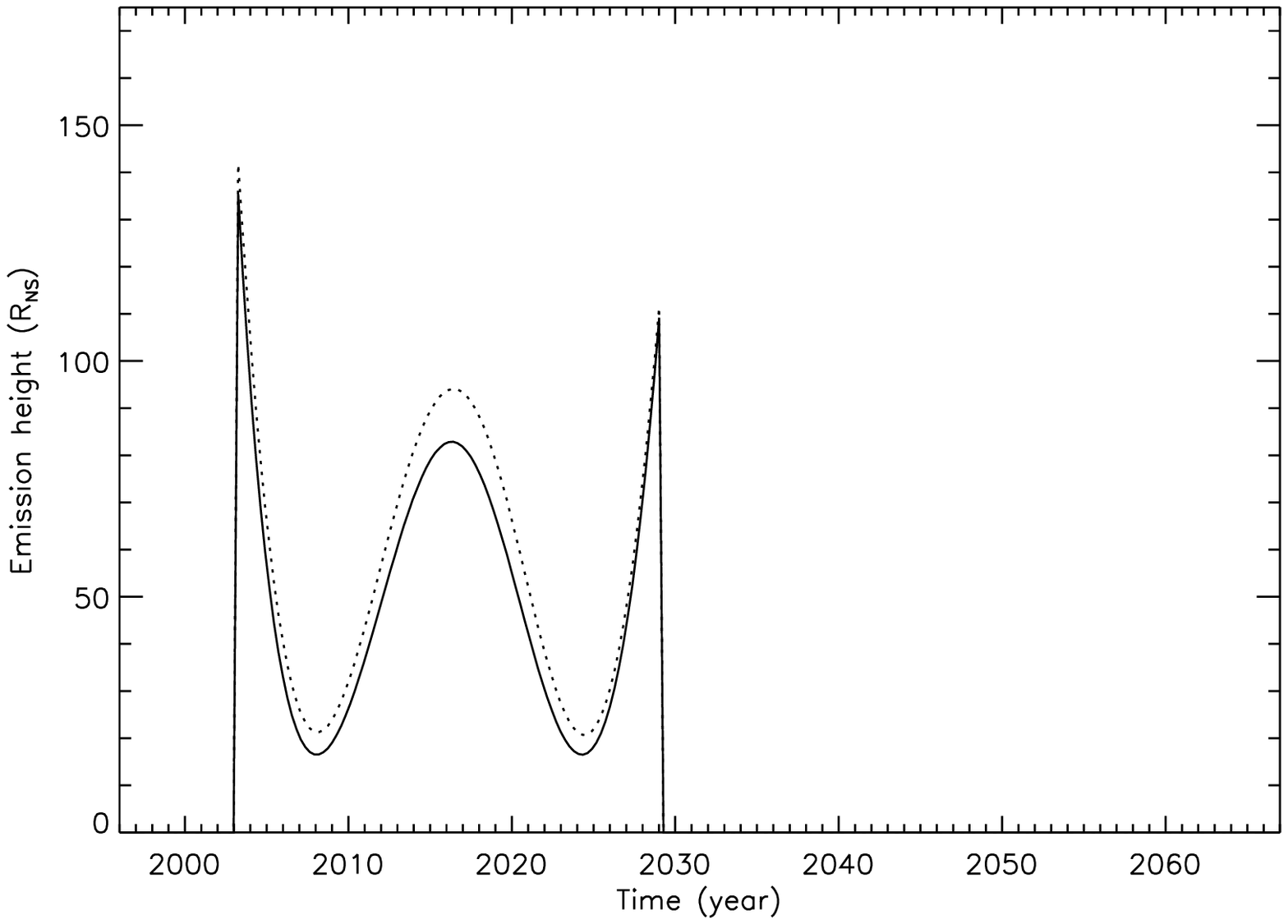}
\caption{
As described in Figure~\ref{mjd_BP1}, but for BP2 (orbital longitude is $290\degr$). Note that in this bright phase, the emission height difference between the leading and trailing components of the beam is not as significant as in BP1.
\label{mjd_BP2}}
\end{figure*}

Also, due to precession of the spin axis, the emission height varies with time because the angle $\beta(t)$ varies with time. Again, there are two different heights for the leading and trailing edges of the beam. These are shown in Figure~\ref{mjd_BP1} and \ref{mjd_BP2}. In BP2, the difference between the two heights is not as significant as in BP1 because of the orbital position of B in BP2. In this region, pulsar B, A and our line-of-sight are roughly aligned, resulting a less deviation in emission heights for the two edges of the beam.

If the emission is produced from the boundary between the open and closed field lines, we can consider these estimates to be the actual emission heights for B. If the emission is produced elsewhere within the open field line region, these are lower limits on emission heights.

\section{Upper limit of the emission height}
\label{uplimit}

We may also set an upper limit to the radio emission height by modeling the distortions of B's magnetosphere induced by the wind of A. Since both the magnetosphere and the wind are strongly magnetized, the distortions depend on the relative strengths of the magnetic fields and thus on the distance from the neutron star. Depending on the location of the radio emission region and the line of sight (and hence on the orbital position) an observer will detect different radiation signatures of the distorted magnetosphere. Inversely, by studying the orbital modulation and using a model of the distorted magnetosphere, we can deduce the location of the emission region.

Similar to how the Sun distorts the Earth's magnetosphere, pulsar A produces a strong enough wind to interact with the magnetic field of B and shape its magnetosphere. The nature of this interaction will vary depending on the properties of the wind. For a wind with a substantial particle flux, the formation of a bow shock, similar to case of the Earth, is expected. In this MHD confinement model, the shape of the Earth's magnetosphere is mostly determined by the pressure balance between the supersonic solar wind and the Earth's nearly dipolar field. This curved shape of the magnetosphere is reproduced well by current numerical models \citep{tsy02a,tsy02b,tsy07}.

On the other hand, for a strongly magnetized wind, reconnection between the wind and the companion's magnetic field lines must be considered. This results in an open structure for the whole magnetosphere, similar to the one originally proposed by \citet{dun61} for planetary magnetospheres.

In the case of the double pulsar, it is unclear whether a MHD confinement model or a reconnection model is more applicable due to the unknown composition of A's wind. However, we are mostly interested in the overall geometric structure of B's magnetosphere. For this purpose, it is sufficient to discuss magnetospheric structure in the most basic terms, relying on the models of planetary magnetospheres. We consider two extreme, though complimentary, models of the Earth's magnetosphere; the highly resistive, analytical reconnection model of \citet{dun61}, hereafter D61, and the fully screened, 3D numerical hydrodynamic confinement model of \citet{tsy02a,tsy02b}, hereafter TS02.

We use the D61 model of the open planetary magnetosphere as a simple analytical representation of the distorted
magnetosphere. The D61 model states that the interplanetary magnetic field (IMF) may become reconnected with the terrestrial field along the day-side magnetopause, where the magnetopause is the boundary between the Earth's magnetosphere and the solar wind. This results in a distortion of the higher altitude regions of the inner magnetosphere.

\begin{figure*}
\epsscale{0.3}
\includegraphics[angle=0,width=16cm]{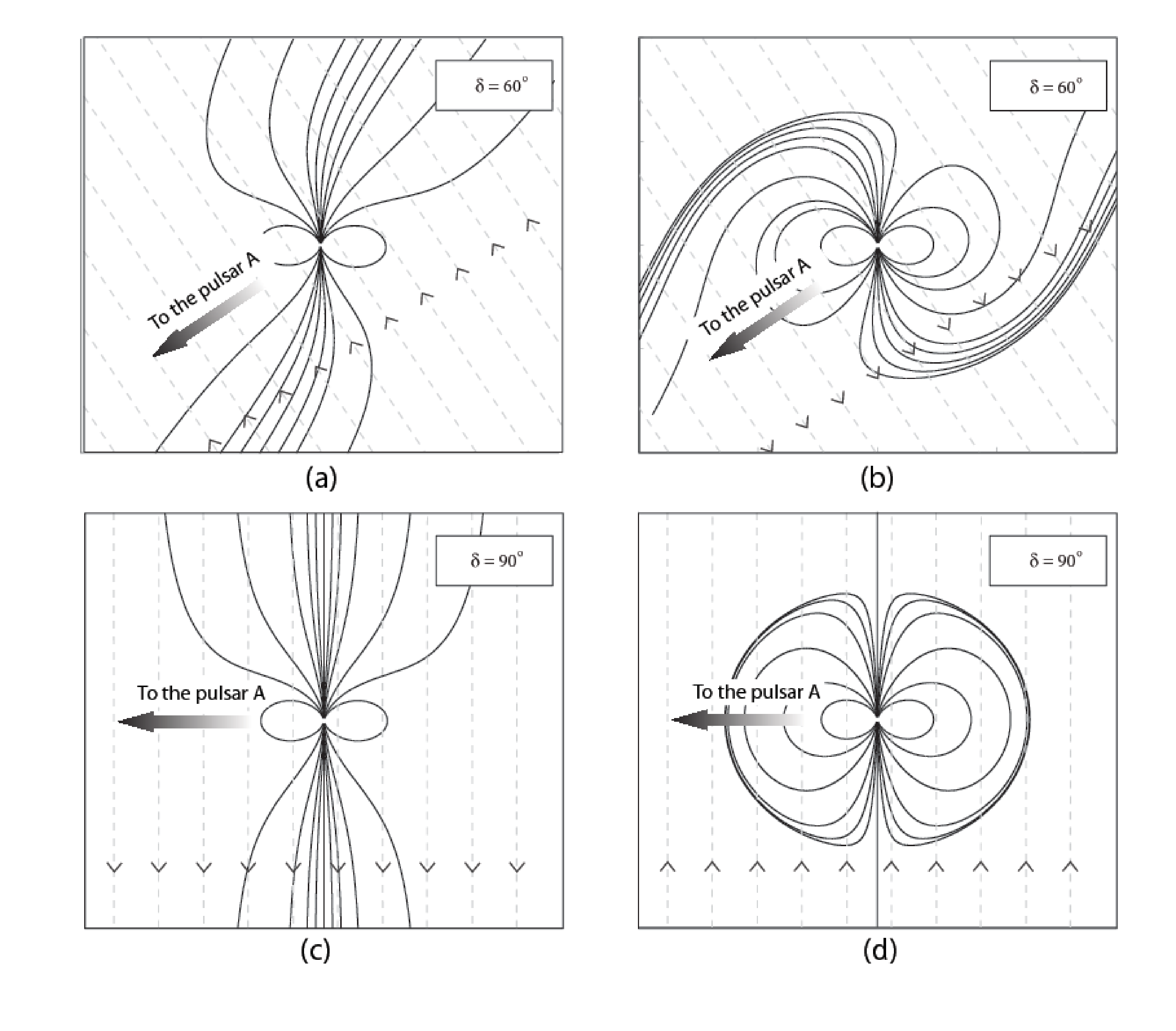}
\caption{
Geometric structures of Dungey-type magnetospheres in 2D. Magnetospheric models constructed by adding a wind's uniform field, in the direction shown by the arrows on the dashed lines, to a dipole field with northward orientation. Magnetic fields of $6.4\times 10^{11}$~G and $10$~G are assumed for the surface magnetic field of pulsar B and the wind magnetic field, respectively. The direction of the wind is the opposite of the arrow showing the direction to the pulsar A. Both (a) and (b) have the same direction of the wind and orientation of the dipole. Same is true for (c) and (d). However, two different possible orientations of the magnetic field in the wind result in very different overall magnetospheric structures. (a) shows smaller deflection of the polar field lines compared to (b). Magnetosphere in (c) is mostly open, whereas one in (d) is mostly closed. In the latter, a radius of the enclosed magnetosphere is about $4\times 10^9$~cm.
\label{magnetosph_dungey}}
\end{figure*}

\citet{fs71} neglected the dynamics of the reconnection processes and modeled the Earth's magnetosphere as a linear superposition of two magnetic fields: the Earth's closed field and the solar wind's uniform field. Following this approach, we can represent B's magnetosphere as a simple addition of the pulsar's dipole field and the wind's uniform field. Similar to the IMF in the D61 model, we treat the magnetic field in A's wind as homogeneous in the vicinity of B, with the direction of the magnetic flux density vector perpendicular to the line connecting the two pulsars (assuming a toroidal field). However, depending on whether the large-scale toroidal field is prograde or retrograde with respect to the orbit, the geometric structure of the magnetosphere can be significantly different (see Figure \ref{magnetosph_dungey}).

Alternatively, we can use TS02 for more precise, three-dimensional modeling. This model is a data-based best-fit representation for the Earth's screened magnetosphere based on a large number of satellite observations. The model provides the option of adding the contributions from external magnetospheric sources such as the ring current, magnetotail current system, magnetopause currents, and the large-scale system of field-aligned currents to the Earth's dipole field.

We used the GEOPACK code repository developed by Tsyganenko, with modifications to match the properties of the double pulsar system. Instead of analyzing every current component in the TS02 model separately, we manipulated the global input parameters of the code which define the geometric structure of the magnetosphere. The shape and scale of the magnetosphere is controlled by the solar wind ram pressure and the dipole tilt only. Variations in the value of the ram pressure change the magnetosphere self-similarly. In the numerical model, the ram pressure is represented by the parameter PARMOD(1) and has units in nPa. PARMOD(2) represents the disturbance storm time (Dst) index and has units in nT. The Dst index is a measure of the size and strength of the ring current, which contributes to the overall field configuration in the inner magnetosphere. The TS02 model is designed in such a way that the structure of the magnetosphere within a stand-off distance from the Earth has a very small dependence on the components of the IMF. Hence, for simplicity we set the transverse components of the external field (PARMOD(3)=$B_{y}$ and PARMOD(4)=$B_{z}$) to zero.

\begin{figure*}
\epsscale{1.8}
\includegraphics[angle=0,width=16cm]{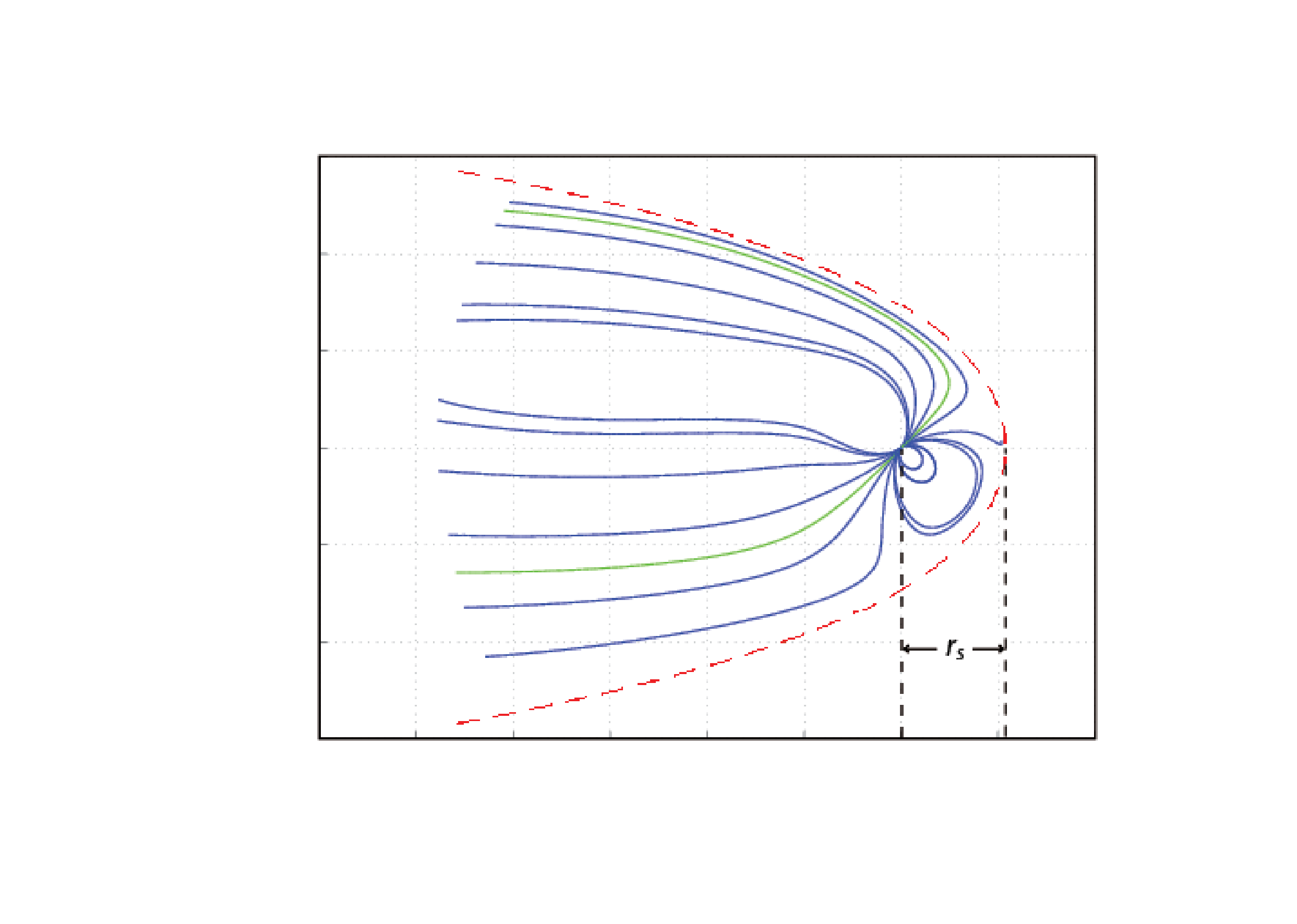}
\caption{TS02 magnetosphere fitted to the theoretical boundary model. Field lines are plotted in solid. Out of which the green color is for the polar field lines. Dashed line represents the bow shock boundary model. Tilt of the dipole is $45\degr$ and $r_{s} \sim 4 \times 10^9$~cm. We fix the shape and scale of the boundary and fit the parameters of the TS02 model.
\label{magnetosph_TS}}
\end{figure*}

We performed a visual fitting (see Figure \ref{magnetosph_TS}) of the boundary produced by the TS02 code to the boundary produced by our theoretical model (equation~(\ref{pressure}) and (\ref{lumB}) in section~\ref{bmodel}). We set PARMOD(3) and PARMOD(4) equal to zero and changed PARMOD(1) and PARMOD(2) until the shapes of the boundaries matched. The best fit values that we obtained are PARMOD(1)=8 nPa for the solar wind ram pressure, PARMOD(2)=100 nT for the Dst index, and, by default, the zero transverse components of the IMF (PARMOD(3)=0 nT, PARMOD(4)=0 nT). This set of parameters produces a magnetosphere boundary with a stand-off distance of about 10.4 stellar radii. In order to make the spatial scaling consistent with the properties of the double pulsar, we rescaled the stellar radius parameter $R0$ from 1 to 0.0026. This change simply ensures that the stand-off distance is about 4000 stellar radii, which is the value assumed throughout this section.

The obtained values of the parameters (PARMOD(1-4) and $R0$) are not supposed to be physically realistic; rather, they produce a magnetosphere with a shape and size (defined by the stand-off distance) that match the properties of the double pulsar. Moreover, there could be other successful fits since they are derived from the visual inspection of the boundaries (see Figure \ref{magnetosph_TS}). Nevertheless, using this particular set suits our purpose of modeling an approximate structure of pulsar B's distorted magnetosphere without using large computational resources.

\begin{figure*}
\epsscale{1.50}
\center
\includegraphics[angle=0,width=13cm]{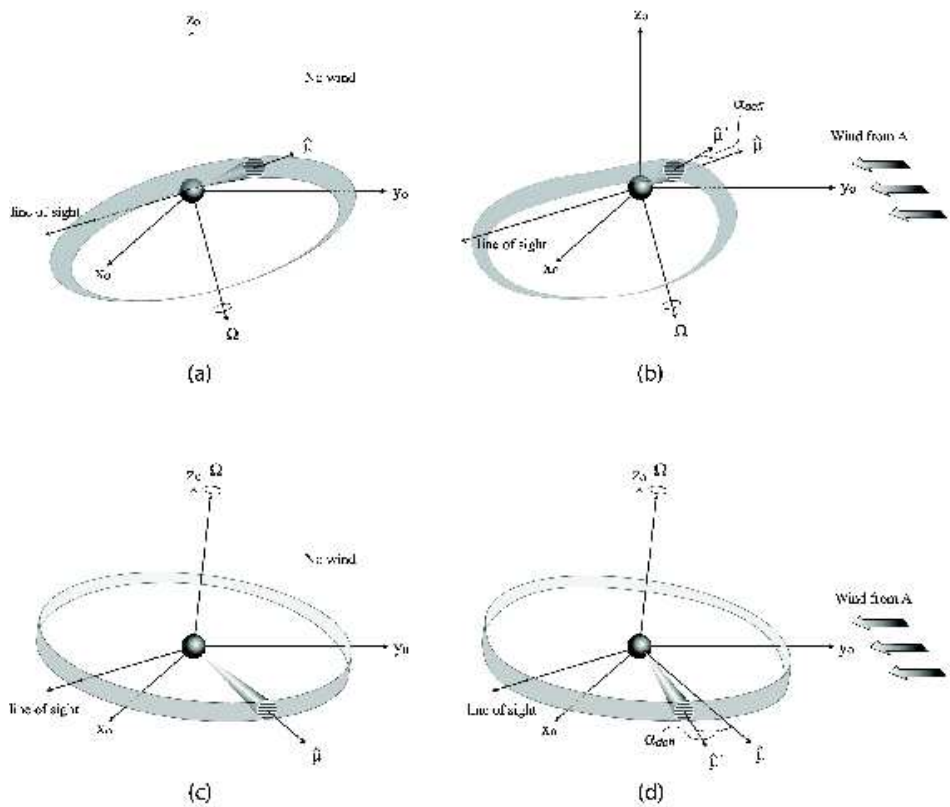}
\caption{Schematic view of the deflection of the emission direction by the wind. (a) and (c) show the configuration of the main axes of an isolated pulsar, with different orientation of the spin axis, however. (b) and (d) show the same picture with an addition of the wind from the companion, for the same configuration of axes as (a) and (c) respectively. Grey-shaded ribbon-type surfaces are the traces of the cross-section of the emission beam. If the line-of-sight intersects this surface then the observer detects the radio emission. In case of (a) and (b), the orientation of the line-of-sight with respect to the spin axis and magnetic axis is such that it does not intersect with the grey-shaded surface after distortion, as shown on (b). This is not the case for (c) and (d) where the line-of-sight intersects with the grey-shaded surfaces, even after distortion.}
\label{distortions_criteria}
\end{figure*}

We employed the same criteria to estimate an upper limit for the emission height for both models. We assumed that the elliptical emission beam is located close to the polar field lines, which are nearly aligned with the magnetic axis at $r \ll R_{LC}$, where $R_{LC}$ is the light cylinder radius. The anisotropic distortion of the magnetosphere by the wind changes the location of the polar field line relative to the undistorted magnetic axis (see Figure \ref{distortions_criteria}, (b) and (d)). As a first approximation, the deflection angle can be expressed as $\alpha_{defl} \sim {B_{w}}/{B_{p}}$, where ${B_{w}}$ and ${B_{p}}$ are the magnetic filed of pulsar A's wind and the magnetic field of pulsar B, respectively. Close to the neutron star's surface, the influence of the wind's magnetic field on the overall field structure is negligible. Therefore, $\alpha_{defl} \sim 0$ at the surface and increases outwards as the wind's magnetic field becomes comparable to the pulsar field near the boundary. The ratio between the two fields, and hence the amplitude of the deflection, depends on the distance from the star as well as on $\delta$, the angle between B's magnetic axis and the line connecting the two pulsars.
There is a certain height above which the distortion is strong enough to deflect the polar field line by more than the angular radius of the beam $(\rho_{a,10} \simeq 14.3\degr)$, which is determined in section~\ref{remodel}. Furthermore, if the component of the distortion perpendicular to the trajectory of the center of the beam in the vicinity of our line-of-sight is large enough, then the emission beam can be pushed away from the line-of-sight to the extent that they do not intersect with each other for any spin phase (see Figure \ref{distortions_criteria}, (b)). This will render the emission beam unseen. On the other hand, the opposite can be true if the distortion occurs mostly along the local trajectory of the beam. In this case, the visibility of the beam stays unchanged and a small shift in the spin phase, at which the emission beam is seen, might be the only observable imprint of the distortions (Figure \ref{distortions_criteria}, (d)). However, the geometry of pulsar B suggests that the former must be realized (Figure \ref{distortions_criteria}, (a) and (b)). Therefore, in order to be able to detect pulsed radio emission from B, the deflection angle $\alpha_{defl}$ should not exceed $14.3\degr$. This places an upper limit on the height of the emission region. However, in order to be able to use this reasoning, a pulsar must be detectable through its pulsed radio emission. We therefore restrict our analysis to  only BP1 and BP2,  the distinct radio-loud regions of the orbit.

In order to make use of this criteria, we analyzed the moments of the closest approach at the orbital phases within BP1 and BP2. At the moment of the closest approach, the magnetic axis is nearly aligned with the line-of-sight, making the angle $\delta$ about the same as the angle between the line-of-sight and the line connecting the two pulsars. In turn, the latter is related to the orbital phase as $(90\degr - \phi_{orb})$(mod $180$). Therefore, $\delta_{*} \sim (90\degr-\phi_{orb})$ when $-90\degr \leq \phi_{orb} < 90\degr$ and $\delta_{*} \sim (\phi_{orb}-90\degr)$ when $90\degr \leq \phi_{orb} < 270\degr$, where $\delta_{*}$ is the value of $\delta$ at the moment of the closest approach. For instance, at the orbital phase $185\degr$, $\delta_{*} \sim 95\degr$ and at $305\degr$, $\delta_{*} \sim 145\degr$. Therefore, $\delta_{*}$ varies within $[95\degr, 145\degr]$ and $[145\degr, 180\degr]$ for BP1 and BP2 respectively.

We calculated the deflection angle $\alpha_{defl}$ using two different methods. We used a simple analytical estimation for the simplified D61 model, while employing more complex numerical calculations for the modified TS02 model. In the D61 model, the system is characterized by three main parameters: the magnetic moment of the pulsar, the  magnetic field in A's wind $\vec{B}_{w} $, and the  angle between the two. Neither of the models depend on the absolute values of the pulsar and wind's magnetic fields. In both cases, only the ratio of these two fields matter. This boils down to the assumption that the wind from A is strongly magnetized (i.e., the shape of the boundary and stand-off distance is defined solely by the magnetic pressure balance). The pulsar wind is highly magnetized near the light cylinder. The particle component only takes over much further, closer to the termination shock. Since the wind from A reaches B after only ~1000 light cylinder radii, the assumption about its high magnetization is valid. Therefore, we can describe both fields by one dimensionless parameter: stand-off distance normalized to the stellar radius. At the moment of the closest approach, the magnetic moment of B is almost aligned with the line of sight. In turn, due to the peculiarity of the double pulsar,  the line of sight is nearly parallel to the orbital plane. Thus, at the moment of the closest approach, B's magnetic axis and the magnetic field of A's wind are nearly coplanar. Therefore, for approximate estimates, the full three-dimensional analysis of the system is not necessary and we only carry out the calculations for the two-dimensional configuration.

In 2D polar coordinates $(r,\lambda)$, the equation for the magnetic field lines reads as

\begin{equation}
\label{deflection1}
\frac{dr}{B_{r}^{tot}}=\frac{r d \lambda}{B_{\lambda}^{tot}}
\end{equation}

\noindent
where $\lambda$ is the colatitude and is equal to $90\degr$ at the equator and to zero along the magnetic axis, which is the same colatitude angle that defined in equation (\ref{dipole}). The angle between the local tangent to the field line and the vector $\vec{r}$ can be approximated as $r d \lambda/dr$. In order to find the deflection angle of the polar field line due to the distortions by the wind, we consider the change in $r d \lambda/dr$.

\begin{equation}
\label{deflection2}
\alpha_{defl}=\left(\frac{r d \lambda}{dr}\right)_{distorted}-\left(\frac{r d \lambda}{dr}\right)_{undistorted}.
\end{equation}

\noindent
As a superposition of the pulsar's dipolar and wind magnetic fields we take a simple addition of the two. Therefore, it follows from the equation~(\ref{deflection1}) that

\begin{equation}
\label{deflection3}
\alpha_{defl}=\left(\frac{B_{\lambda}^{p}+B_{\lambda}^{w}}{B_{r}^{p}+B_{r}^{w}}\right)-\left(\frac{B_{\lambda}^{p}}{B_{r}^{p}}\right).
\end{equation}

\noindent
Here, $B_{r}^{p}$,$B_{\lambda}^{p}$ and $B_{r}^{w}$, $B_{\lambda}^{w}$ are $r$ and $\lambda$ components of the pulsar and wind magnetic fields, respectively. The angle between the local components of the fields can be either $\delta_{*} + \lambda + 90\degr$ or $90\degr - (\delta_{*} + \lambda)$, depending on the orientation of the toroidal field. Thus, in the frame of the dipole, the magnetic field components can be expressed as follows

\begin{eqnarray}
&& B_{r}^{p}= -\frac{\mu_{0} m }{2 \pi} \frac{\cos \lambda}{r^{3}} \label{deflection41}\\
&& B_{\lambda}^{p}=\frac{\mu_{0} m }{4 \pi} \frac{\sin \lambda}{r^{3}} \label{deflection42}\\
&& B_{r}^{w}=\mp B_{0}^{w} \sin(\lambda+\delta_{*}) \label{deflection43}\\
&& B_{\lambda}^{w}=\mp B_{0}^{w} \cos(\lambda+\delta_{*}). \label{deflection44}
\end{eqnarray}

\noindent
Here, $m$ is a magnetic moment of pulsar B while $B_{0}^{w}$ is a strength of the wind magnetic field. The wind magnetic field structure is believed to be toroidal. However, for the sake of simplicity, we assumed a locally uniform wind field across the whole magnetosphere of B, which is feasible since the radius of the light cylinder is much smaller than the orbital radius. In the equations (\ref{deflection43}) and (\ref{deflection44}), $B_{r}^{w}$ and $B_{\lambda}^{w}$ can switch signs depending on whether the large scale toroidal field of the wind is prograde with respect to the orbital motion of the pulsars or retrograde. Mathematically it is equivalent to replacing $\delta_{*}$ with $\delta_{*} + 180\degr$. Below, we derive the approximate expression for $\alpha_{defl}$ for the upper signs in the equations (\ref{deflection43}) and (\ref{deflection44}) and only in the end substitute $\delta_{*} + 180\degr$ instead of $\delta_{*}$ to account for both cases.

It is reasonable for our case $(\alpha_{defl} \leq 14.3\degr)$ to limit our estimations to the field lines close to the magnetic axis, i.e. $\lambda \sim 0\degr$. Then we can rewrite the equations (\ref{deflection41})-(\ref{deflection44}) in the following way

\begin{eqnarray}
&& B_{r}^{p}= -\frac{\mu_{0} m }{2 \pi} \frac{1}{r^{3}} \label{deflection81}\\
&& B_{\lambda}^{p}=0 \label{deflection82}\\
&& B_{r}^{w}= -B_{0}^{w} \sin \delta_{*} \label{deflection83}\\
&& B_{\lambda}^{w}= -B_{0}^{w} \cos \delta_{*}. \label{deflection84}
\end{eqnarray}

\noindent
Here, we are only with left the upper signs in the second pair of equations.
We get an approximate expression for the deflection angle by substituting (\ref{deflection81})-(\ref{deflection84}) into (\ref{deflection3})

\begin{equation}
\label{deflection9}
\alpha_{defl} = \frac{ B_{0}^{w} \cos \delta_{*}}{\frac{\mu_{0} m }{2 \pi} \frac{1}{r^{3}} + B_{0}^{w} \sin \delta_{*}}.
\end{equation}

\noindent
We can rewrite equation (\ref{deflection9}) in terms of normalized distance $\bar{r} \equiv r / r_{s}$, where $r_{s}=\left(\mu_{0} m / 2 \pi B_{0}^{w}\right)^{1/3}$ is a stand-off distance ($r_{s} \sim 4 \times 10^9$~cm as estimated from our boundary model),

\begin{equation}
\label{deflection10}
\alpha_{defl} = \frac{ \bar{r}^{3} \cos \delta_{*}}{1 + \bar{r}^{3} \sin \delta_{*}}.
\end{equation}

\noindent
From the criteria for the pulsed radio emission detectability, it follows that the values of $\bar{r}$ for which $\alpha_{defl} > 14.3\degr$ must be excluded as possible emission heights. To find such values of $\bar{r}$ for any $\delta_{*}$, we use the condition that the absolute value of the right-hand side of the equation (\ref{deflection10}) must exceed $14.3\degr$. Hence, in the case of a prograde toroidal field we have

\begin{equation}
\label{deflection11}
\left| \frac{\bar{r}^{3} \cos \delta_{*}}{1 + \bar{r}^{3} \sin \delta_{*}} \right| > 14.3\degr.
\end{equation}

\noindent
By replacing $\delta_{*}$ with $\delta_{*}+180\degr$, we get the detectability criteria for the retrograde configuration

\begin{equation}
\label{deflection12}
\left| \frac{- \bar{r}^{3} \cos \delta_{*}}{1 - \bar{r}^{3} \sin \delta_{*}} \right| > 14.3\degr.
\end{equation}

\begin{figure*}
\epsscale{1.8}
\center
\includegraphics[angle=-90,width=13cm]{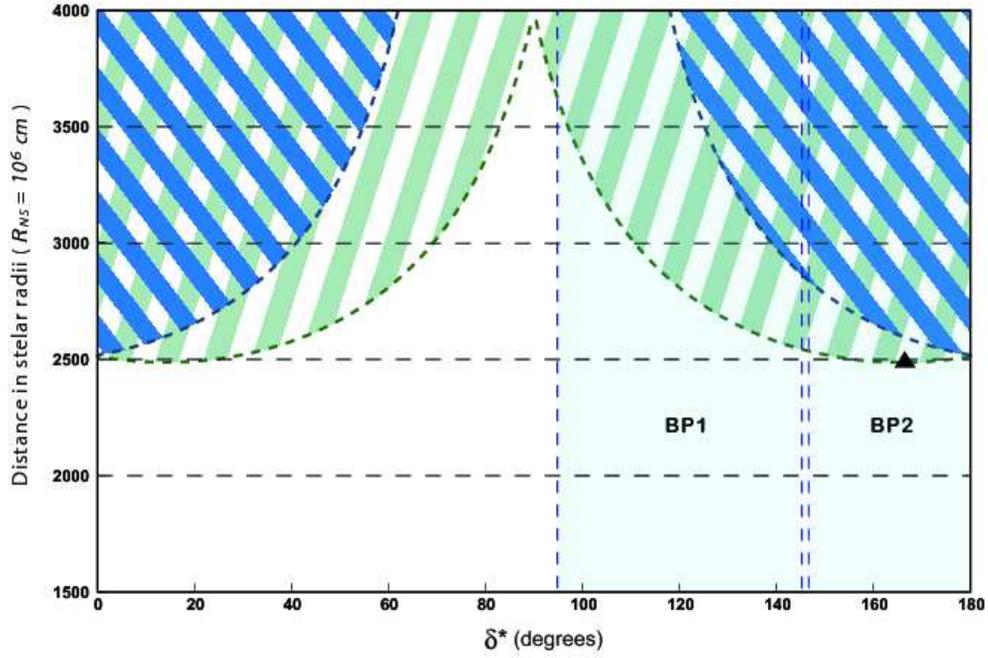}
\caption{Excluded values of the radio emission height calculated by using the analytical D61 model. The curtain-shaped shaded areas represent the domains for emission height and $\delta^{*}$ for which the deflection angle exceeds $14.3\degr$. The area with \emph{blue} stripes leaned to the left corresponds to the prograde configuration (equation~\ref{deflection11}), while the area with \emph{green} stripes leaned to the right corresponds to the retrograde configuration (equation~\ref{deflection12}). The shaded areas between the vertical dashed lines represent the values of $\delta^{*}$ valid for the analysis (i.e. those corresponding to BP1 and BP2). The black triangle marks the minimum value of $\sim 2500 NS$ radii within the shaded range of $\delta^{*}$. Thus, $2500 NS$ radii is the best upper limit for the emission height within the D61 model.}
\label{D61model}
\end{figure*}

\noindent
The minimum of the values of $\bar{r}$ that satisfy equations (\ref{deflection11}) or (\ref{deflection12}) for any $\delta_{*}$ corresponding to BP1 and BP2 is the best upper limit we can put on the emission height with this approach. In Figure~\ref{D61model}, we show the solutions of the equation (\ref{deflection11}) and (\ref{deflection12}), represented by the shaded areas over the contours of constant deflection angle of $14.3\degr$. As we can see in Figure \ref{D61model}, both orientations of the wind magnetic field produce almost the same upper limit, which is $\sim 2500 R_{NS}$. This value corresponds to $\delta_{*} \sim 170\degr - 180\degr$, i.e. when pulsar B, pulsar A and the Earth are nearly aligned, where pulsar B is in BP2 region. This is consistent with the method by which we estimate an upper limit. At superior conjunction, the wind magnetic field is perpendicular to the magnetic axis, resulting in the largest distortion of the polar field lines. Therefore, the distance from the star surface, above which the deflection is more than $14.3\degr$, is smallest at superior conjunction.

We can use the same criteria to set an upper limit on the emission height using the numerical TS02 model. We require that the numerically calculated distortion angle of the polar field line must not exceed the angular radius of the beam $(\rho_{a,10} \simeq 14.3\degr)$. This allows us to find the maximum emission height for each value of $\delta_{*}$.

\begin{figure*}
\epsscale{1.80}
\includegraphics[angle=-90,width=16cm]{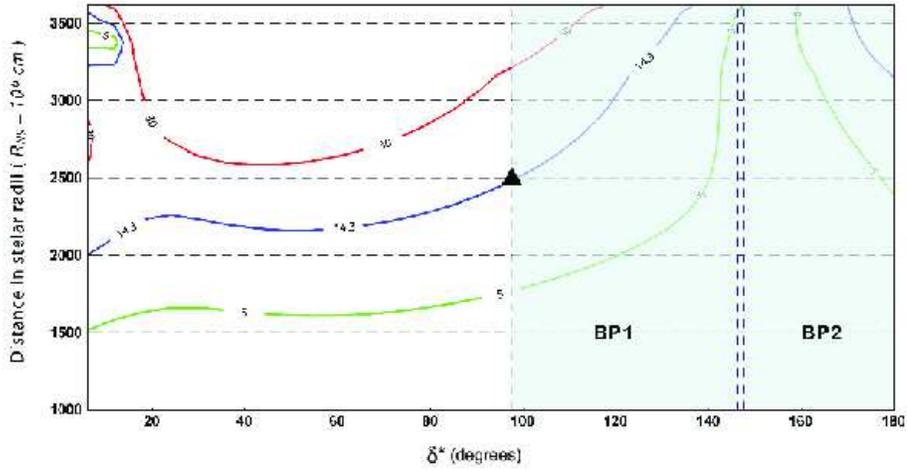}
\caption{Fixed value contours for the deflection angle calculated by using the modified TS02 model. Each contour shows the upper limits to the emission region height for the corresponding value of the deflection angle and $\delta^{*}$. The shaded area between the vertical dashed lines represents the values of $\delta^{*}$ valid for the analysis (i.e. those corresponding to BP1 and BP2). The black triangle on the $14.3\degr$ contour marks the minimum value of 2500 NS radii within the shaded range of $\delta^{*}$. Thus, 2500 NS radii is the best
upper limit for the emission height within the modified TS02 model.}
\label{TS02model}
\end{figure*}

Using the modified TS02 code, we can trace any field line (particularly polar field lines) of B's distorted magnetosphere for any orientation of the magnetic axis with respect to A's wind. This means that we can calculate the deflection angle for any $\delta_{*}$ at any altitude. The $14.3$ contour on Figure~\ref{TS02model} shows the altitudes at which the deflection angle equals $14.3\degr$ for all values of $\delta_{*}$. The altitude with the lowest value amongst others is the best upper limit we can put on the emission height. Since we can only consider $\delta_{*}$'s corresponding to BP1 and BP2, the resultant best upper limit of the emission height would be $2500 R_{NS}$ for $\delta_{*} \sim 95\degr$ (orbital phase of $185\degr$, which is in BP1)(see Figure \ref{TS02model}).

In summary, we adapted the models of the Earth's distorted magnetosphere to the double pulsar system, based on the similarities between the two. We adjusted the spin-orbital and magnetic field parameters corresponding to the observational data. Both magnetospheric models, the analytic D61 and numerical TS02, draw simplified and very extreme pictures of the double pulsar. Nevertheless, both models offer an improvement over a simple dipole and allow us to set an upper limit on the altitude of the emission region using a novel technique. Moreover, they would greatly compliment each other if somehow unified into one model.

We used the criteria of the pulsar emission detectability to estimate an upper limit of the emission height. This requires the distortion angle of the polar field line not to exceed the angular radius of the beam, derived from the observational data.

Moreover, we arrived at similar results by employing two very different models; the analytical, highly resistive Dungey type model and the numerical, fully screened modified TS02 model. Both approaches led to the conclusion that B's radio emission is generated within the inner $22\%$ of the light cylinder.

\section{Discussion}
\label{dis}

The determination of  radio pulsar emission heights is important for understanding their emission mechanisms. Pulsar B of the double pulsar system provides a unique opportunity to study different emission regions of the magnetosphere due to precession. Also, the magnetosphere is distorted, exhibiting a complicated field line structure, due to the wind of A.
These distortions depend on the orbital and rotational phases of B. Observations of these distortions, not observed in isolated pulsars, via the  orbital variations of the radio intensity of B allow us to pinpoint the location of radio emission.

We have applied a simple wind-magnetosphere interaction model to determine the boundary of the magnetosphere of  B. The best solution describes the shape of the boundary as a polynomial, with coefficients dependent on the angle between the magnetic axis and the line connecting the two pulsars. Furthermore, the boundary was not axially symmetric, but for simplicity we modified it to be symmetric. The stand-off distance ranges from $3.8-4.5\times10^{9}$~cm according to the orientation of the magnetic axis with respect to the line connecting two pulsars, resulting in a size of the polynomial boundary of less than 30\% of the light cylinder radius. However, this size is three times larger than the size inferred from eclipses of  A. Thus, a possible explanation for this is that the particle density in the magnetosphere of B falls off significantly as a function of the radial distance from the center of the pulsar, so that the radiation of  A penetrates the outer regions of B's magnetosphere.

Moreover, the variation of the boundary will change the shape of the open field line region. As a result, the spin-down luminosity of  B can vary slightly due to the variation in the area of the polar cap. This causes a $1.5\%$ periodic variation in the spin-down luminosity. It can also lead to a correction on the spin phase, but this is very small \citep{gll+11}. Thus we did not consider this effect in our model.

As we determined, the range of the allowed emission height depends on the orbital motion due to the relative orientation of the magnetic axis with respect to the boundary. Also,  precession changes the location of the spin axis, so that the emission height changes with time. In both of these variations, we have been calculating the emission heights for both the leading and the trailing edges of the conal elliptical beam. For a normal pulsar with its light-cylinder boundary, these two edges give the same height due to  cylindrical symmetry. When the impact parameter is equal to the angular radius of the beam across the semi-major axis, we would detect a single-peak profile, resulting in one emission height. Figure~\ref{orb} shows that the relative heights of emission due to the two components of the beam switch in the two bright phase regions since the orientation of the boundary changes throughout the orbit. These estimated emission heights are about 1\% of the light cylinder radius or 4\% of the stand-off distance. Again, these should be considered lower limits if the emission does not originate on the last closed field line.
Moreover, the analytical and numerical approaches to the upper limit estimate lead to the conclusion that pulsar B's radio emission is generated within $22\%$ of the light cylinder.

In normal pulsars, radio emission heights have been calculated by using their geometry and the pulse profile widths \citep{kg97}. These range from about $10$ to $100~R_{NS}$, less than 10\% of the light cylinder radius. Our emission height estimates are consistent with these results. Thus the radio emission produced by B likely has the same mechanism as for isolated pulsars, which is consistent with \citet{lyu05}.
Most theories of pulsar radio emission place the generation region close to the star, typically within one stellar radius \citep[e.g.][]{mel95}. In contrast, a model based on the anomalous cyclotron-Cherenkov resonance \citep{mu79,lbm99} requires emission to be generated much higher up in the magnetosphere, at hundreds of stellar radii. The fairly high emission altitudes of radio emission inferred in the present paper are consistent with the latter models.

The magnetospheres of pulsars can be distorted due to rotation as proposed in \citet{dh04}, resulting in a rotational sweepback of the magnetic field lines. They found that at low altitude the rotation deflects the local direction of the magnetic field line by at most an angle of the order of $(r/R_{LC})^2$, where $r$ is the radial distance of the field line. We applied this rotational sweepback model to pulsar B along with our boundary model and found that the rotational sweepback is very small, because the deflection of the magnetic field line from its local direction is of order $0.1$ radians. The estimated upper limit for the emission height implies that this effect is less than $0.05$ radians, negligible compared to the distortions by the wind. However, this effect is significant when the radial distance of the field is close to the light cylinder, so that it is important in normal pulsars.

The relativistic phase-shift method can be used to determine the radio emission heights of pulsars as described in \citet{gg01} and \citet{drh04}. This method uses aberration and retardation effects to explain the observed pulse phase shift of  pulse profiles containing  core and conal components. We applied this method to the double-peaked pulse profiles of pulsar B to estimate the emission height. Because pulsar B has only a conal component, we assumed that pulse phase of zero was at the minimum between the two peaks. Then the phase-shift is measured from the two peaks, leading and trailing,  with respect to this reference phase. The calculation shows that the phase-shift method does not work for pulsar B. For example, the emission height on MJDs 53860 and 53939 is 6 and 23 NS radii, respectively. On MJD 54050 it is zero due to zero phase shift. Also on MJD 54400, the phase of trailing component is larger than the absolute phase of leading component, so that the emission height becomes negative. The reason of these calculated height fluctuations is that the pulse profile of B is not stable and varies significantly. Therefore, it is difficult to measure the shift in pulse phase accurately. Also, as there are only two peaks in the pulse profile of B, the determination of the pulse phase zero reference point is difficult. Therefore, the measured shifts and then the emission heights may not be correct, concluding that this method cannot be used to constrain emission heights of pulsar B. However, this is a useful method of estimating emission heights of normal pulsars which have stable pulse profiles with both core and conal emission components \citep{gg03}.

In summary, by using the method presented in this paper, we can place limits on  the radio emission height for any pulsar with well-determined emission geometry. The advantage of this method is that by estimating the field-line constant by tracing the magnetic field lines, we can constrain the emission heights of pulsars which have high magnetic inclinations that make them unsuitable for the other methods. Our radio emission height estimations for pulsar B will be useful for future studies and in particular can be used to constrain proposed geometrical models such as \citet{lyu05} and \citet{fwk+09} in order to accurately explain the observations.

\acknowledgments

BBPP is supported by NRAO student observing support and MAM is supported by a WV EPSCoR grant. The National Radio Astronomy Observatory is a facility of the National Science Foundation operated under cooperative agreement by Associated Universities, Inc.

\bibliographystyle{apj.bst}
\bibliography{psrrefs,modrefs,journals}

\end{document}